\documentclass[12pt]{article}

\input epsf
\usepackage{amssymb,amsmath,wrapfig,subfigure}
\usepackage[matrix,arrow,curve]{xy}
\usepackage{cite}
\usepackage{graphicx,color}
\usepackage[debug,pageanchor=false]{hyperref}
\definecolor{link}{rgb}{.8,.15,.1}
\hypersetup{colorlinks=true,linkcolor=link,citecolor=link,urlcolor=link,linktocpage}

\makeatletter
\@addtoreset{equation}{section}
\makeatother

\def\rr {{\Bbb R}}

\def\zz {{\Bbb Z}}
\def\del {\partial}

\def\del {\partial}
\def\tts {{$T\oplus T^*$}}
\def\stt {{$\mathrm{SU(3)}\times\mathrm{SU(3)}$}}
\def\sut {{${\rm SU}(3)$}}

\begin{document}

		       \begin{titlepage}
		       \begin{center}

		       \vskip .3in \noindent

%		       {\Large \bf{Supersymmetry of the O6-plane with Romans mass}}
		       {\Large \bf{Localized O6-plane solutions with Romans mass}}
		       \bigskip

			 Fabio Saracco and Alessandro Tomasiello\\

		       \bigskip
				 Dipartimento di Fisica, Universit\`a di Milano-Bicocca, I-20126 Milano, Italy\\
		       and\\
		       INFN, sezione di Milano-Bicocca,
		       I-20126 Milano, Italy
		       %\bigskip

		       \vskip .5in
		       {\bf Abstract }
		       \vskip .1in

		       \end{center}
		       %\vskip .4in

		       %\noindent
		\hyphenation{dif-fer-en-tial}

		Orientifold solutions have an unphysical region around their source; for the O6, the singularity is resolved in M-theory by the Atiyah--Hitchin metric. Massive IIA, however, does not admit an eleven-dimensional lift, and one wonders what happens to the O6 there. In this paper, we find evidence for the existence of localized (unsmeared) O6 solutions in presence of Romans mass, in the context of four-dimensional compactifications. As a first step, we show that for generic supersymmetric compactifications, the Bianchi identity for the $F_4$ RR field follows from constancy of $F_0$. Using this, we find a procedure to deform any O6--D6 Minkowski compactification at first order in $F_0$. For a single O6, some of the symmetries of the massless solution are broken, but what is left is still enough to obtain a system of ODEs with as many variables as equations. Numerical analysis indicates that Romans mass makes the unphysical region disappear.
		
		       \vfill
		       \eject

		       \end{titlepage}	

	\hypersetup{pageanchor=true}
	% to fix a bug with titlepage&hyperref: see comment after \usepackage[debug,pageanchor=false]{hyperref}

	       \tableofcontents

\section{Introduction} % (fold)
\label{sec:intro}

For a long time, string theory was unsatisfactory in that its definition was only perturbative in the string coupling $g_s$. Fortunately, the duality revolution of the mid-'90s allowed to understand the strong coupling limit of almost all perturbative definitions of string theory. In particular, the strong-coupling limit of IIA was realized to be eleven-dimensional supergravity. 

Actually, there is a fly in the ointment, which is the reason we wrote	``almost''. Whenever the so-called Romans mass $F_0$ is present, the duality between IIA and eleven-dimensional supergravity does not work any more. Moreover, since $F_0$ is simply one of the field strengths in the theory, this ``massive'' version of IIA can be connected to the ordinary, ``massless'' version by domain walls: the D8-branes.

In \cite{ajtz}, it was shown that, for classical solutions of massive IIA, the string coupling is bounded by the curvature in string units. In a sense, this makes the problem more rare: any solution with large $g_s$ is already invalidated by being strongly curved. This makes the need for a non-perturbative completion less pressing. 

There do exist, however, solutions for whose existence we have independent arguments, which are less easy to dismiss in spite of being strongly curved. One example is the supergravity solution for the D8-brane itself. In this solution, the string coupling on the brane is a free parameter, and it can be made large. This is not in contradiction with the general argument in \cite{ajtz}, because at the same time the solution becomes strongly curved. But in this case we do expect the solution to survive in fully-fledged string theory, because of its open-string interpretation and because it is half-BPS. Similar considerations apply to other configurations such as D4--D8 systems. 

For D6-branes and O6-planes, however, the situation is less clear. 
O6-planes are of particular theoretical interest because of the way they get resolved by M-theory. The metric in the massless theory reads
\begin{equation}\label{eq:o6metric}
	ds^2_{\rm O6} = Z^{-1/2} dx^2_\parallel + Z^{1/2} dx^2_\perp \ ,\qquad
	Z= 1 - \frac{r_0}r \ ,\qquad r_0 = l_s g_s \ .
\end{equation} 
Even if we excise the unphysical ``hole'' $r < r_0$, the metric becomes singular for $r\to r_0$. Of course, probing the metric at such small distances was unwise to begin with: if $g_s$ is small, $r_0$ is smaller than the string length $l_s$. If $g_s$ is large, we may try and use M-theory, with the customary formula $ds^2_{11}= e^{-2 \phi/3}(ds^2_{10}+ e^{2 \phi} (d z + A)^2)$ for the eleven-dimensional lift (and knowing that the dilaton is given by $e^\phi= g_s Z^{-3/4}$). The metric $ds^2_{11}$ is Ricci-flat, but it is still singular at $r=r_0$. Quantum effects, however, correct this metric at small $r$ to the so-called Atiyah--Hitchin metric \cite{atiyah-hitchin,gibbons-manton-AH}. This metric still has a minimum allowed value for the radial variable, but it is now at $r=\frac\pi2 r_0$, and the geometry there is smooth. 

So the singularity of the massless O6 solution (\ref{eq:o6metric}) is resolved in M-theory to a smooth hole. What about O6-planes in \emph{massive} IIA? Solutions of this type have been assumed to exist, especially in the context of flux compactifications. A popular trick in supergravity is to ``smear'' sources over the internal manifold; namely, to replace the localized source with one which is spread all over space. For an orientifold plane in string theory, this is not really physically allowed, since such sources are supposed to sit on the fixed loci of the orientifold involutions. Nevertheless, smeared solutions are often a good indicator of whether a bona fide background will exist. Using this sleight of hand, quite a few massive O6 solutions have been found. A well-known early example \cite{dewolfe-giryavets-kachru-taylor,acharya-benini-valandro}\footnote{The localization of smeared sources was analyzed in \cite{blaback-danielsson-junghans-vanriet-wrase-zagermann}, in different setups from the one we consider here.} of moduli stabilization is of this type. Also, the presence of both O6's and $F_0$ is considered the most promising avenue for producing de Sitter vacua in string theory which are completely classical (as opposed to de Sitter vacua such as \cite{kklt,balasubramanian-berglund-conlon-quevedo}); examples with the smearing trick include \cite{silverstein-simple,danielsson-haque-koerber-shiu-vanriet-wrase}.

It would be interesting, then, to check whether such massive O6-planes really do exist as localized solutions, and if so, what happens to the unphysical hole around their source. 

In this paper, we will find evidence for the existence of \emph{supersymmetric} massive O6-plane solutions. We will mostly consider a spacetime of the form
\begin{equation}\label{eq:ads4}
	{\rm AdS}_4\times M_6\ ,
\end{equation}
since we have already at least the example \cite{dewolfe-giryavets-kachru-taylor,acharya-benini-valandro}, which is of this form. The O6 will be filling the four-dimensional spacetime, as well as three of the six directions in $M_6$. We will also consider the possibility ${\rm Mink}_4\times M_6$; however, we do not know of any supersymmetric Minkowski compactification with O6-planes and Romans mass, and for this reason we will give more attention to (\ref{eq:ads4}). 

Actually, although some of our considerations will be more general, as we get more concrete we will focus on what happens close to the O6 source, so that we can forget about the details of the internal topology; in practice, this just means taking $M_6 = \rr^6$. We cannot expect the geometry on this $\rr^6$ to approach flat space, however, as would be the case if one factorized the metric (\ref{eq:o6metric}) as ${\rm Mink}_4\times \rr^6$. This is because neither ${\rm AdS}_4\times \rr^6$ nor ${\rm Mink}_4\times \rr^6$ are vacua for the massive theory. So taking $M_6=\rr^6$ just means that we are focusing on what happens close to the O6 source; there is really no such thing as a massive O6 solution `in flat space'. Still, one can arrange so that the deviations from flat asymptotics happen at large distances. We are introducing two new length scales: $\frac1{\sqrt{-\Lambda}}$ and $\frac1{g_s F_0}$ (since $F_0$ always appears multiplied by $e^\phi$ in the equations of motion). When both of these scales are large, it is possible to study the features of the geometry closer to the source (order $r_0=g_s l_s$).

\bigskip

Let us now summarize our results, and give a synopsis of the paper. In section \ref{sec:susy}, we give a very condensed review of the ``generalized complex geometry'' approach to supersymmetry, which we will use in the rest of the paper. This formalism divides all possible supersymmetric vacua into three classes, of which only two are relevant here: SU(3) structure, and \stt\ structure. The massless O6 solution, whose metric we gave in (\ref{eq:o6metric}) and which we review at greater length in section \ref{sec:o6}, is of the first type; so is the smeared massive O6 solution of \cite{dewolfe-giryavets-kachru-taylor,acharya-benini-valandro}, which we review in section \ref{sec:dgkt}. A very simple argument, however, shows that the localized (that is, unsmeared) massive O6 solution we are looking for should be of \stt\ structure type. 

Hence, in section \ref{sec:stt}, we apply the generalized complex formalism of section \ref{sec:susy} to the \stt\ structure case. We find handier expressions for the RR fluxes than were available so far; this allows us to show that the Bianchi identity for $F_4$ follows from that of $F_0$ (which simply requires $F_0$ to be constant). This turns out to be very useful later on; in particular, it is crucial for our first-order deformation in section \ref{sec:def}. This is a simple procedure to deform any SU(3) structure Minkowski solution into an AdS solution, at first order in
\begin{equation}\label{eq:mu}
	\mu \equiv \sqrt{-\frac\Lambda3}
\end{equation}
(where $\Lambda$ is the cosmological constant). 

Any Minkowski solution obtained as back-reaction of an O6--D6 system on a Calabi--Yau is of SU(3) structure type\footnote{In fact, conversely, all \emph{known} smeared Minkowski solutions can be obtained as O6--D6 systems on Calabi--Yau's, up to duality.}. So we can apply the first-order deformation procedure to any such configuration. To go beyond first-order, however, we need to have more symmetries, and for this reason in section \ref{sec:massiveo6} we focus on the neighborhood of an O6 source, in the sense explained earlier. 

Actually we find (in section \ref{sub:symmetries}) that already the first-order solution has a smaller symmetry group than the massless O6 solution. This smaller group is fortunately still large enough to reduce the problem to a system of ODEs. As we show in section \ref{sub:higher}, this turns out to have as many variables as equations (thanks in part to the result in section \ref{sec:stt} about the Bianchi identity of $F_4$), and is thus expected to admit a solution. We performed both a perturbative and a numerical study. The latter seems to work well enough to infer some properties of the solution.

In particular, we find that the unphysical ``hole'' around the orientifold source gets resolved by the Romans mass; see figure \ref{fig:o6num}. For $\mu\to0$, $F_0\to 0$, we recover the massless O6 solution (\ref{eq:o6metric}), as we should; raising both these quantities in an appropriate way, the unphysical region disappears, while the string coupling and the curvature stay small. 

For the theoretical status of massive IIA, this is a very satisfying outcome. Massless O6 solutions have an unphysical region around their sources, which gets resolved by M-theory; in the massive theory, there is no lift to M-theory, but massive O6 solutions do not have the unphysical region to begin with. 

The $S^2$ that surrounds the O6 does not shrink as $r \to 0$: it remains of finite size. It is possible, however, to continue the metric analytically with essentially another copy of the same geometry. 

% section intro (end)

\section{Supersymmetry} % (fold)
\label{sec:susy}

We will begin by reviewing in this section the conditions for unbroken supersymmetry for four-dimensional compactifications, using the language of generalized complex geometry.

\subsection{The equations in general} % (fold)
\label{sub:susy}

We will consider a spacetime of the warped-product form
\begin{equation}\label{eq:warped}
	ds^2_{10}= e^{2A}ds^2_4 + ds^2_6\ ,
\end{equation}
where $ds^2_4$ is the metric for either Mink$_4$ or AdS$_4$, $ds^2_6$ is the metric on the internal manifold $M_6$ (which is otherwise left arbitrary) and $A$ is a function of $M_6$ called \emph{warping}.
 
Such a geometry is supersymmetric in type IIA\footnote{The conditions for type IIB, that we do not need here, are obtained by $\phi_+\leftrightarrow \phi_-$.} if and only if \cite[Sec.7]{gmpt3}
\begin{itemize}
    \item There exists an \stt\ structure $\phi_\pm$ on $M_6$.
    Here, $\phi_\pm$ are polyforms which are pure spinors for Clifford$(6,6)$, and which satisfy
\begin{equation}\label{eq:stt}
    (\phi_+,\bar\phi_+)=(\phi_-,\bar \phi_-)\neq 0
    \ ,\qquad
    (\phi_+, X \cdot \phi_-)=0=(\phi_+, X\cdot \bar\phi_-)
\end{equation}
for any $X \in $ \tts. We have used the Chevalley--Mukai internal product
between internal forms: $(\omega,\omega')\equiv (\omega\wedge \lambda(\omega'))_6$,
$\lambda(\omega)\equiv (-1)^{\frac12{\rm Int}({\rm deg}(\omega))}\omega$.
    \item
    There exist a closed three-form $H$, and an even-degree polyform $F = \sum_k F_{2k}$ (the sum of all the {\it internal} fluxes) such that \cite{gmpt2,t-reform}
\begin{subequations}\label{eq:psp}
	\begin{equation}\label{eq:pspM}
	    d_H \phi_+ =0 \ ,\qquad
		d_H (e^{-A} {\rm Re} \phi_-)=0
	    \ ,\qquad
	    {\cal J}_+\cdot d_H(e^{-3A} {\rm Im}  \phi_-)=
	      F
	    \ ,\qquad d_H F= \delta \ 
	\end{equation}
	in the Minkowski case, and \cite{gaiotto-t2} 
	\begin{equation}\label{eq:pspA}
	    d_H \phi_+ = -2 \mu\, e^{-A} {\rm Re} \phi_-
	    \ ,\qquad
	    {\cal J}_+\cdot d_H(e^{-3A} {\rm Im}  \phi_-)=
	    -5 \mu e^{-4A} \,{\rm Re} \phi_+
	    +  F
	    \ ,\qquad d_H F= \delta \ 
	\end{equation}	
\end{subequations}
in the AdS case.
Here, $\Lambda= -3 \mu^2$ is the cosmological constant (as we already saw in (\ref{eq:mu})), and
$d_H \equiv (d - H\wedge)$. $\delta$ is a delta-like source supported on branes or orientifolds. ${\cal J}_+\cdot$ is an algebraic operator that depends on $\phi_+$ alone; it is reviewed for example in \cite[Sec.~2.1]{t-reform}. In section \ref{sub:alg} we will give its explicit expression for the cases we are interested in.
\end{itemize}

Neither the metric $g$, nor the dilaton $\phi$, nor the spinorial parameters $\eta^{1,2}$ of the supersymmetry transformations, appear directly in (\ref{eq:psp}). Moreover, we will soon see that the $H$ which does appear is not actually the physical NS three-form. Rather, there is a one-to-one correspondence between solutions $\phi_\pm$ of the algebraic conditions (\ref{eq:stt}) and the geometric data we just mentioned:
\begin{equation}\label{eq:phigbp}
	\phi_\pm \ \leftrightarrow (g,b,\phi,\eta^{1,2}_\pm)\ .
\end{equation}
In particular, we will call
\begin{equation}
	b_{\phi_\pm}
\end{equation}
the two-form $b$ determined by (\ref{eq:phigbp}). For more details on this correspondence, see \cite{gualtieri,gmpt3}. These data are not really needed to solve (\ref{eq:psp}). However, for completeness we will give in section \ref{sub:alg} the explicit formulas for $(g,b,\phi)$ in terms of $\phi_\pm$ for the cases that we are interested in. 

(\ref{eq:psp}) is invariant under the transformation\footnote{This property is the main reason we are using the system (\ref{eq:psp}) rather than the original form of these equations \cite{gmpt2,gmpt3}, involving the Hodge star. Those equations can be made invariant under (\ref{eq:btrans}) only after defining a rather awkward $*_b \equiv e^b * e^b$ operator.}
\begin{equation}\label{eq:btrans}
	H \to H - d \delta b \ ,\qquad F \to e^{-\delta b\wedge} F \ ,\qquad
	\phi_\pm \to e^{-\delta b\wedge} \phi_\pm \ .
\end{equation}
As it turns out, the $b_{\phi_\pm}$ determined by $\phi_\pm$ via (\ref{eq:phigbp}) transforms as $b_{\phi_\pm}\to b_{\phi_\pm} + \delta b$ under (\ref{eq:btrans}). The physical NS three-form is the combination
\begin{equation}\label{eq:Hphysgen}
	H_{\rm phys} = H + d b_{\phi_\pm} \ ,
\end{equation}
which is thus invariant under (\ref{eq:btrans}). The physical RR field is the one which obeys $d_{H_{\rm phys}} F_{\rm phys}= \delta$:
\begin{equation}
	F_{\rm phys}= e^{b_{\phi_\pm}\wedge} F\ .
\end{equation}

% subsection susy (end)

\subsection{Solving the algebraic constraints} % (fold)
\label{sub:alg}

We will now analyze the algebraic part of the supersymmetry equations,
(\ref{eq:stt}).

There are three cases to consider. Let us call
the {\it type} of a pure spinor $\phi= \sum_{k\ge k_0} \phi_k$ the smallest degree $k_0$ that appears in the sum; in other words, $\phi$
only contains forms of degree type$(\phi)$ or higher. It turns out that the type of a pure spinor in dimension 6 can be at most 3.
There are then three cases:
\begin{enumerate}
    \item $\phi_+$ has type 0, and $\phi_-$ has type 3. This is usually referred to as the ``\sut\ structure'' case, for reasons that will become clear soon.
    \item $\phi_+$ has type 0, and $\phi_-$ has type 1. This is the most generic case, and for this reason it is sometimes just called ``\stt'', or also ``intermediate SU(2) structure''.
    \item $\phi_+$ has type 2, and $\phi_+$ has type 1. This is called ``static SU(2) structure'' case.
\end{enumerate}

In this paper, we will only need the first two cases. 

\subsubsection{${\rm SU}(3)$} % (fold)
\label{ssub:su3}

We will start by giving the solution of the algebraic constraints in (\ref{eq:stt}) in the SU(3) structure case. The condition of purity on each $\phi_\pm$ separately determines (up to a $b$-transform)
\begin{equation}\label{eq:su3}
    \phi_+= \rho e^{i \theta} e^{-iJ} \ ,\qquad \phi_- = \rho \,\Omega\ ,
\end{equation}
with $\rho$ and $\theta$ real functions, $J$ a non-degenerate real two-form, and $\Omega$ a decomposable three-form (one that can be locally written as wedge of three one-forms) such that $\Omega\wedge \bar\Omega$ is never zero\footnote{We are including $(\phi,\bar \phi)\neq 0$ in the definition of purity.}.
The constraint (\ref{eq:stt}) then reduces easily to
\begin{equation}\label{eq:compsu3}
    J \wedge \Omega = 0 \ ,\qquad
    J^3= \frac 34 i \Omega\wedge \bar \Omega\neq 0 \ .
\end{equation}
These equations define an SU(3) structure, which justifies the name given earlier to case 1. 

We will now describe the map (\ref{eq:phigbp}) for this case. The $b_{\phi_\pm}$ obtained by it is zero:
\begin{equation}
	b_{\phi_\pm}= 0 \ .
\end{equation}
The metric defined by $\phi_\pm$ (which, once again, is not needed in the system (\ref{eq:psp})) can be described as follows. $\Omega$, being decomposable, determines an almost complex structure $I$ (it is the one such that $\Omega$ is a $(3,0)$-form).
Then we write $g= J I$. The condition (\ref{eq:compsu3}) implies that the $g$ defined in this way is symmetric. Finally, the dilaton is given by
\begin{equation}\label{eq:dilSU3}
    e^{\phi} = \frac{e^{3A}}{\rho}\ .
\end{equation}

We also give the form of ${\cal J}_+\cdot$, which enters (\ref{eq:psp}):
\begin{equation}\label{eq:Jdotsu3}
	{\cal J}_+ = J\wedge - J^{-1}\llcorner \ .
\end{equation}

% subsubsection su3 (end)

\subsubsection{\stt} % (fold)
\label{ssub:su3su3}

In this case, one can parameterize the most general solution to (\ref{eq:stt}) as \cite{jeschek-witt,minasian-petrini-zaffaroni, halmagyi-t} 
\begin{subequations}\label{eq:nicepair}
    \begin{align}\label{eq:nice+}
        \phi_+ = &
		\rho\, e^{i \theta}
		    \exp[ -i J_\psi] 
		\ ,\\
        \label{eq:nice-}
        \phi_- = &\rho\, v\wedge \exp[i \omega_\psi]\ ,
    \end{align}
\end{subequations}
where
\begin{equation}\label{eq:Jpsi}
	J_\psi \equiv \frac1{\cos(\psi)}j + \frac i{2 \tan^2(\psi)} v\wedge \bar v \ ,\qquad 
	\omega_\psi \equiv \frac1{\sin(\psi)} \left({\rm Re} \omega + \frac i{\cos(\psi)} {\rm Im} \omega\right)\ ,
\end{equation}
for some (varying) angle $\psi$, real function $\rho$, one-form $v$ and two-forms $\omega, j$ satisfying
\begin{equation}\label{eq:SU2}
    \omega^2=0 \ ,\qquad \omega \wedge \bar \omega = 2 j^2 \ ,
\end{equation}
which mean that $\omega, j$ define an SU(2) structure.\footnote{Actually, from the constraint (\ref{eq:stt}), one would get (\ref{eq:SU2}) wedged with $v\wedge \bar v$, but one can show \cite[Sec.~3.2]{halmagyi-t} that these can be dropped without any loss of generality.} These can also be rewritten more symmetrically as 
\begin{subequations}\label{eq:SU2q}
	\begin{align}
		&j \wedge {\rm Re} \omega = {\rm Re} \omega \wedge {\rm Im} \omega = {\rm Im} \omega \wedge j = 0\ , \label{eq:jRo} \\
		&j^2=({\rm Re} \omega)^2 = ({\rm Im} \omega)^2\ ; \label{eq:j2}
	\end{align}
\end{subequations}
these equations are reminiscent of the defining relations of the quaternions $i,j,k$, which is ultimately because SU(2)$\cong$Sp(1). Finally, the inequality in (\ref{eq:stt}) implies that the top-form $v\wedge \bar v \wedge j^2$ should be non-zero everywhere.

We will now detail the map (\ref{eq:phigbp}) for this case. This can be inferred by comparing (\ref{eq:nicepair}) to its derivation in \cite{jeschek-witt,minasian-petrini-zaffaroni, halmagyi-t} from spinor bilinears. For example, \cite[Eq.~(3.19)]{halmagyi-t} can be connected to (\ref{eq:nicepair}) by a $b$-transform; from this, we see that the $b_{\phi_\pm}$ defined by (\ref{eq:phigbp}) is non-zero:
\begin{equation}\label{eq:b}
    b_{\phi_\pm}= \tan (\psi) {\rm Im}  \omega\ .
\end{equation}

The metric can be found by relating the forms $j$, $\omega$ and $v$ in (\ref{eq:nicepair}) to the spinor bilinears of an SU(3) structure. In \cite{halmagyi-t} one finds $J=j + \frac i2 z\wedge \bar z$, $\Omega= \omega\wedge z$, where $z\equiv \frac1{\tan (\psi)} v$. This tells us that the metric is the direct sum of a two-by-two block $z\bar z= \frac1{\tan^2(\psi)} v \bar v$, and of a four-by-four block determined by the SU(2) structure $j$, $\omega$. In other words, we have two orthogonal distributions (namely, subbundles of $T$): $D_2$ and $D_4$. The explicit form of the four-by-four block in the metric is $g_4 = j I_4$, where $I_4$ is an almost complex structure along $D_4$. This means that $I_4$ squares to -1 along $D_4$: 
\begin{equation}
	I_4^2 = - \Pi_4 \ ,\qquad
\end{equation}
where $(\Pi_4)^m{}_n= \delta^m{}_n - {\rm Re} v^m {\rm Re} v_n - {\rm Im} v^m {\rm Im} v_n$ is the projector on $D_4$. We should now compute $I_4$.  This can be done by writing $I_4 = ({\rm Re} \omega)^{-1} {\rm Im}  \omega$ (which can be derived in holomorphic indices). Since $\omega$ only spans four directions, ${\rm Re} \omega$ has rank 4; so writing $({\rm Re} \omega)^{-1}$ is an abuse of notation. It should be understood as an inverse along the distribution $D_4$. In practice, it can be computed as a matrix of minors: 
\begin{equation}
	[({\rm Re} \omega)^{-1}]^{mn} = -2 \frac{(dx^m \wedge dx^n \wedge {\rm Re}  \omega \wedge v \wedge \bar v)}{({\rm Re} \omega)^2 \wedge v \wedge \bar v}\ .
\end{equation}
Putting all together, we have 
\begin{equation}\label{eq:metric}
    ds^2 = j I_4  +
    \frac1{\tan^2 (\psi)} v \bar v \ , 
 	\ ,\qquad I_4=({\rm Re} \omega)^{-1} {\rm Im} \omega \ .
\end{equation}

Finally, the dilaton $\phi$ is determined by\footnote{This corrects Eq.~(3.15) in \cite{gaiotto-t2}. We thank A.~Zaffaroni for helping us find this mistake.}
\begin{equation}\label{eq:rho}
    e^{\phi} = \frac{e^{3A}}{\rho}\cos(\psi)
\end{equation}
for both cases considered in this subsection, (\ref{eq:su3}) and (\ref{eq:nicepair}).

We also give the form of the operator ${\cal J}_+ \cdot$ that appears in (\ref{eq:psp}) is similar to the one in (\ref{eq:Jdotsu3}): 
\begin{equation}\label{eq:Jdot}
	{\cal J}_+ = J_\psi\wedge - J_\psi^{-1}\llcorner \ .
\end{equation}

% subsubsection su3su3 (end)

% subsection alg (end)

% section stt (end)

\section{O6 solution} % (fold)
\label{sec:o6}

In this brief section, we will review how the O6 solution in flat space, whose metric was given in (\ref{eq:o6metric}), solves the system (\ref{eq:pspM}). 

The internal space $M_6$ is in this case nothing but 
$\rr^6$, with coordinates $x^i$ and $y^i$ (to be thought of
respectively as parallel and orthogonal to the O6).

The O6 solution is of ${\rm SU}(3)$-structure type (\ref{eq:su3}). For cosmological constant $\Lambda = 0$, and hence $\mu = \sqrt{-\Lambda/3}=0 $, the equations in (\ref{eq:pspM}) read
\begin{align}\label{eq:rho1}
	&\rho=e^{3A-\phi}={\rm const} \ ,\qquad d J = 0 = H \ ; \qquad
	d ( e^{-A} {\rm Re} \Omega) = 0 \\ 
	& F_2 = -J^{-1}\llcorner d(e^{-\phi} {\rm Im} \Omega)
	\ ,\qquad d F_2 = \delta\ . \label{eq:F2su3}
\end{align}
Notice that, in this case, $\theta$ is constant, but otherwise undetermined. 

In general, in (\ref{eq:rho1}) $\delta$ is a delta-like current supported on the sources present. For the O6 solution, it reads
\begin{equation}\label{eq:deltaO6}
	\delta=\delta_{\rm O6}= - 4\pi l_s\delta(y^1)\delta(y^2)\delta(y^3) 
	dy^1 \wedge dy^2 \wedge dy^3\ ;
\end{equation}
an SU(3) structure that solves (\ref{eq:rho1}) can then be given as
\begin{equation}\label{eq:JOO6}
	J= dx^i\wedge dy^i\ , \qquad \Omega=
	i(Z^{-1/4}dx^1 +i Z^{1/4}dy^1 )\wedge
	(Z^{-1/4}dx^2 +i Z^{1/4}dy^2 )\wedge
	(Z^{-1/4}dx^3 +i Z^{1/4}dy^3 )\ 
\end{equation}
with $Z$ the Green function for the flat Laplacian in $\rr^3$:
\begin{equation}\label{eq:Z}
 	Z = 1-\frac {r_0}{r} \ ,\qquad r\equiv \sqrt{y^i y^i} \ ,\qquad
	r_0 = g_s l_s \ ,
\end{equation}
as we already saw in (\ref{eq:o6metric}).\footnote{If we had had $N$ D6-branes instead of an O6-plane, the function $Z$ would have read $1+\frac{r_0}{r}$, with $r_0=N l_s g_s/2$.}
 We also have
\begin{equation}\label{eq:F2O6}
	F_2 
	= -\frac {l_s}{2r^3} \epsilon_{ijk} y^i dy^j \wedge dy^k
	\ ,\qquad e^A=Z^{-1/4} \ ,\qquad e^\phi = g_s Z^{-3/4} \qquad \left(\rho= \frac1{g_s}\right)\ ;
\end{equation}
$g_s$ is a constant that we can think of as the value of $e^\phi$ at infinity. 

The SU(3) structure in (\ref{eq:JOO6}) is one possible solution to (\ref{eq:rho1}), and by itself it only describes four supercharges; there are other solutions, related to the one in (\ref{eq:JOO6}) by flipping some signs, which describe the other supercharges. In this paper, we will focus on (\ref{eq:JOO6}) and ignore the SU(3) structures: for this reason, our massive solutions will have ${\cal N}=1$ supersymmetry.

Finally, notice that, since the solution stops making sense before we can get to $r\to 0$, the equation $d F_2 = \delta$ has to be understood as a Gauss' law: namely, 
\begin{equation}
	\int_{S^2} F_2 = -4\pi l_s \ ,
\end{equation}
for any $S^2$ that surrounds the origin, where the O6-plane is located.

% section o6 (end)

\section{Smeared O6 with Romans mass} % (fold)
\label{sec:dgkt}

Our aim is to find a O6 solution in the presence of Romans mass. As recalled in the introduction, a solution of this type can be found easily if one ``smears'' the O6 source; this was done in \cite{dewolfe-giryavets-kachru-taylor} in the language of effective field theory, and lifted to ten dimensions in \cite{acharya-benini-valandro}. 

We take a spacetime of the form (\ref{eq:ads4}): the four-dimensional part has non-zero cosmological constant. This means that $\mu \neq 0$, and thus we have to use the version (\ref{eq:pspA}) of the supersymmetry conditions. If we also take 
\begin{equation}\label{eq:th0}
	\theta = 0 \ ,
\end{equation}
we get\footnote{The first two equations in (\ref{eq:pspA}), which are the ones that are equivalent to the conditions of unbroken supersymmetry, do not by themselves imply that $A=0$. For the Romans mass they would give $g_s F_0 = 5 \mu e^{4A}$; if one now also adds the Bianchi condition $dF_0=0$, one gets that $A$ is constant. In (\ref{eq:SU3th0}) we set it to zero, because a non-zero value can always be reabsorbed in the definition of $\mu$.}
\begin{equation}\label{eq:SU3th0}
	\begin{split}
		 &dJ=0 \ ,\qquad d \Omega= -i g_s F_2\wedge J \ ,\qquad 
		H= 2 \mu {\rm Re} \Omega \ ,\qquad \rho={\rm const} \ ,\qquad A=0\ ; \\
		&g_s F_0 = 5 \mu \ ,\qquad dF_2 - H F_0 = \delta \ ,\qquad
		g_s F_4 = \frac32 \mu J^2 \ ,\qquad F_6 = 0 \ . 
	\end{split}	
\end{equation}
From (\ref{eq:dilSU3}), it also follows that the dilaton is constant;  $g_s\equiv e^\phi$.

So far, the source $\delta$ was unspecified. To find the solution in \cite{dewolfe-giryavets-kachru-taylor}, take $F_2=0$. Then we see that the Bianchi identity for $F_2$ implies 
\begin{equation}
	\delta= - 2 \mu F_0 {\rm Re} \Omega \ .
\end{equation}
This is the ``smearing'' proposed in \cite{acharya-benini-valandro}. 

To get a sense of the physics of this compactification, let us moreover assume as in \cite{dewolfe-giryavets-kachru-taylor} that $F_0$ is of order one, that the periods of $F_4$ are of order $N$, and that the internal space has volume $\sim R^6$. We know already that $\delta \propto {\rm Re}  \Omega$; it makes sense to fix the proportionality constant as
\begin{equation}\label{eq:smear}
	\delta \sim - \frac 1 {R^3} {\rm Re} \Omega \ ,
\end{equation}
so that integrating $\delta$ along a 3-cycle gives an order one number.
The Bianchi identity then says $F_0 \mu \sim R^{-3}$; moreover, from (\ref{eq:SU3th0}) we see that $F_0 \sim \mu/g_s$ and $F_4 \sim F_0 R^4$. We thus find that the parameters scale as
\begin{equation}\label{eq:scales}
	R\sim N^{1/4}\ ,\qquad g_s \sim \mu \sim R^{-3}\sim N^{-3/4}\ .
\end{equation} 

We have seen that it is easy to find a supersymmetric solution including O6 planes and Romans mass, if one is willing to smear the O6 source $\delta$ as in (\ref{eq:smear}). As stressed in the introduction, smearing an O6 is not really meaningful in string theory, but solutions obtained with this trick are often precursors to ``localized'' solutions, namely ones where the source is delta-like as it should be (as in (\ref{eq:JOO6})). So we can take the solution reviewed in this section as an inspiration for the solution we are looking for. 

The most natural course of action might seem to solve the equations (\ref{eq:SU3th0}) without assuming $F_2=0$, and with an unsmeared source, unlike in (\ref{eq:smear}). However, we immediately face a problem: (\ref{eq:SU3th0}) imposes $A=0$. This does not seem possible for a solution with a source: in particular, the solution with $F_0=0$ has a non-constant $A$, as we can check from (\ref{eq:F2O6}).

So unfortunately we cannot use SU(3) structure solutions. We are left with the cases 2 and 3 in section \ref{sub:alg}. If we think of adding a small amount of $F_0$ to the massless solution, which is SU(3), it seems more natural to select case 2, which is generic and can be continuously connected to the SU(3) structure case, rather than case 3, which is isolated. This is the reason we did not study case 3 in section \ref{sub:alg}. In section \ref{ssub:su3su3} we reviewed the solution (\ref{eq:nicepair}) of the algebraic constraints (\ref{eq:stt}) for case 2; we will now analyze the  corresponding differential equations.

% section dgkt (end)

\section{\stt\ structure compactifications} % (fold)
\label{sec:stt}

As we just saw, a localized O6 with Romans mass cannot be an SU(3) structure solution; this motivates us to look for an \stt\ structure solution. For that class, the algebraic constraints have been reviewed in section \ref{ssub:su3su3}; we will now use those results (in particular (\ref{eq:nicepair})) in the system (\ref{eq:psp}). This section contains both a review of old results, and some new ones --- most importantly, the expressions for the fluxes. 

For reasons explained in the introduction, we will first look at the AdS case, which we divide in two sections, \ref{sub:ads} and \ref{sub:adst0}. We will then also analyze the Minkowski case, in section \ref{sub:mink}.

\subsection{AdS: generic case} % (fold)
\label{sub:ads}

\subsubsection{Geometry} % (fold)
\label{ssub:fluxless}

We will start by the first equation in (\ref{eq:pspA}), $d_H \phi_+ = - 2 \mu e^{-A}{\rm Re} \phi_-$. Using (\ref{eq:nicepair}), the one-form part says that
\begin{align}\label{eq:drho}
    &d(\rho \,\sin(\theta))=0\ , \\ 
\label{eq:rev}    
	&{\rm Re} v = \frac{e^A}{2 \mu \sin(\theta)} d \theta \ .
\end{align}
In deriving (\ref{eq:rev}), we have solved (\ref{eq:drho}) by taking 
\begin{equation}\label{eq:rho0}
	\rho=\frac{\rho_0}{\sin(\theta)}\ ,
\end{equation}
where $\rho_0$ is a constant. This means that we have assumed 
\begin{equation}
	\theta\neq 0\ 
\end{equation}
everywhere. In this subsection, we will continue our analysis in this assumption. The case $\theta=0$ is quite different, and will be described in section \ref{sub:adst0}.

Coming back to $d_H \phi_+ = - 2 \mu e^{-A}{\rm Re} \phi_-$, its three-form part now gives
\begin{align}\label{eq:H}
    &H= -d (\cot(\theta) J_\psi)\ , \\
    \label{eq:dJpsi}
    &d\,\left(\frac1{\sin(\theta)} J_\psi \right) =2\mu e^{-A}{\rm Im} (v\wedge \omega_\psi)%=
    %\frac{ 2 \mu e^{-A}}{\sin(\psi)}\left(
    %{\rm Im}v\wedge {\rm Re} \omega + {\rm Re} v\wedge
    %\frac{{\rm Im} \omega}{\cos(\psi)}\right)
	\ .
\end{align}
Finally, the five-form part can be shown to follow from the
one- and three-form parts, (\ref{eq:rev}) and (\ref{eq:dJpsi}).

% subsubsection fluxless (end)

\subsubsection{Flux} % (fold)
\label{ssub:flux}

We will now look at the second equation in (\ref{eq:pspA}). We have seen that $H$ is determined by (\ref{eq:H}). We can then use (\ref{eq:btrans}) with the choice $\delta b=-\cot(\theta) J_\psi$, so that we end up with $H=0$ in (\ref{eq:pspA}). 

However, there is a price to pay. Once we transform $\phi_+ \to e^{-\delta b\wedge} \phi_+$, we also have to transform the associated operator ${\cal J}_+ \cdot$:
\begin{equation}
	{\cal J}_+ \cdot \to e^{- \delta b\wedge} {\cal J}_+ \cdot e^{\delta b\wedge}\ .
\end{equation}
For the choice $\delta b=-\cot(\theta) J_\psi$, remembering (\ref{eq:Jdot}), we get that the new ${\cal J}_+$ operator is 
\begin{equation}
	{\cal J}_+\cdot = e^{\cot(\theta)J_\psi\wedge} (-J_\psi^{-1}\llcorner + J_\psi\wedge) e^{-\cot(\theta)J_\psi\wedge}\ .
\end{equation}
This can be computed in two ways. The first is to compute the associated action on $T\oplus T^*$, where $e^{b\wedge}$ is represented by $\bigl( \begin{smallmatrix} 1&0\\ -b&1
\end{smallmatrix} \bigr)$. The second is to just use the formula $e^{-A} B e^A= B + [B,A]+\frac12[B,[B,A]]+\ldots$, and  
\begin{equation}
	[J_\psi^{-1}\llcorner,J_\psi\wedge]= h \ ,\qquad h \omega_k\equiv 
	(3-k) \omega_k \ ,
\end{equation}
as an example of the usual Lefschetz representation of ${\rm Sl}(2,\rr)$ on forms (see for example \cite[Ch.~0.7]{griffiths-harris}). Either way, we get
\begin{equation}
	{\cal J}_+\cdot = - J_\psi^{-1}\llcorner + \cot(\theta) h +\frac1{\sin^2(\theta)} J_\psi\wedge \ .
\end{equation}

We can now compute the fluxes from the second equation in (\ref{eq:pspA}):
\begin{subequations}\label{eq:Fk}
	\begin{align}
	    & F_0 = -J_\psi^{-1}\llcorner
	    d (\rho e^{-3A}{\rm Im} v)
	    + 5 \mu \rho e^{-4A}\cos(\theta) \label{eq:F0}\ ;\\
	\label{eq:F2}
		& F_2 = F_0 \cot(\theta) J_\psi -J^{-1}_\psi \llcorner d\, {\rm Re} (\rho e^{-3A} v \wedge \omega_\psi) \\
		\nonumber&\hspace{1cm}+ \mu \rho e^{-4A}
		\left[(5+2\tan^2(\psi)) \sin(\theta) J_\psi +2 \sin(\theta) {\rm Re} v \wedge {\rm Im} v- 2\cos(\theta)\frac{\sin(\psi)}{\cos^3(\psi)} {\rm Im} \omega\right]\ ;
	 \\
	& F_4 = F_0 \frac{J^2_\psi}{2 \sin^2(\theta)} + 
	d\Big[ \rho\, e^{-3A} (J_\psi \wedge {\rm Im} v - \cot(\theta) {\rm Re} (v \wedge\omega_\psi))\Big] \label{eq:F4}\ ;
	\\
	& F_6 = -\frac1{\cos^2(\psi)}{\rm vol}_6 \left(F_0\frac{\cos(\theta)}{\sin^3(\theta)} + 3 \frac{\rho \mu e^{-4A}} {\sin(\theta)}\right)\ .  
	\end{align}
\end{subequations}
Recall that $\rho$ is related to the dilaton by (\ref{eq:rho}). The expression for $F_0$ already appeared in \cite{gaiotto-t2}. The expressions for $F_2$ and $F_4$ are new; their expressions appear much simpler than in earlier computations, thanks in part to the $\delta b$ transformation we performed earlier. 

Notice that the Bianchi identities for (\ref{eq:Fk}) are now $d F_k=0$, away from sources. The one for $F_0$ just says $F_0$ is constant, as usual. If we now consider $dF_4$, we see that the term not multiplying $F_0$ is exact, so it drops out. On the other hand, the form $J_\psi^2/\sin^2(\theta)$ that multiplies $F_0$ is easily seen to be closed as a consequence of (\ref{eq:dJpsi}). So we conclude
\begin{equation}\label{eq:dF4}
	dF_0=0\ \Rightarrow\ dF_4=0\ .
\end{equation} 
In other words, the Bianchi identity for $F_4$ is redundant. This fact will be very important for the rest of this paper. 

We should stress once again that the $F_k$ given in (\ref{eq:Fk}) are the ones which are closed under $d$ --- and which are locally given by $F_k = d C_{k-1}$. The physical NSNS three-form is given by combining (\ref{eq:Hphysgen}), (\ref{eq:b}) and (\ref{eq:H}):
\begin{equation}
	H_{\rm phys} = d B_{\rm phys}= d (-\cot(\theta) J_\psi +\tan(\psi){\rm Im} \omega)\ ;
\end{equation}
the RR fluxes which are closed under $(d- H_{\rm phys}\wedge)$ are then given by
\begin{equation}
	\tilde F = e^{B_{\rm phys}\wedge} F\ . 
\end{equation}

% subsubsection flux (end)

% subsection ads (end)

\subsection{AdS: special case} % (fold)
\label{sub:adst0}

We will again start by the first equation in (\ref{eq:pspA}), $d_H \phi_+ = - 2 \mu e^{-A}{\rm Re} \phi_-$. Our generic analysis in section \ref{sub:ads} relied on the assumption that $\theta\neq0$; in this section we will consider the case 
\begin{equation}
	\theta=0\ .
\end{equation}
This obviously solves (\ref{eq:drho}). The remaining one-form equation now says
\begin{equation}\label{eq:revt0}
	{\rm Re} v= - \frac{e^A d \rho}{2 \mu \rho} \ ,
\end{equation}
which replaces (\ref{eq:rev}).

The three-form part of $d_H \phi_+ = - 2 \mu e^{-A}{\rm Re} \phi_-$ now gives
\begin{equation}\label{eq:Ht0}
	d(\rho J_\psi)=0 \ ,\qquad    
	H= 2\mu e^{-A} {\rm Re} (i v \wedge \omega_\psi) \ .
\end{equation}
Finally, the five-form part can be shown to follow from the
one- and three-form parts, (\ref{eq:revt0}) and (\ref{eq:Ht0}).

We now turn to the RR fluxes. Unlike in section \ref{ssub:flux}, this time there is no natural $b$-transform to perform, because $H$ given in (\ref{eq:Ht0}) is not necessarily exact. So we will give the expressions of the fluxes which are closed under $d_H$, rather than under $d$:
\begin{subequations}\label{eq:Fkt0}
	\begin{align}
	    & F_0 = -J_\psi^{-1}\llcorner
	    d (\rho e^{-3A}{\rm Im} v)
	    + 5 \mu \rho e^{-4A} \label{eq:F0t0}\ ;\\
		& F_2 = -J^{-1}_\psi \llcorner d\, {\rm Im} (i \rho e^{-3A} v \wedge \omega_\psi) - 2 \mu \rho e^{-4A} \frac{\sin(\psi)}{\cos^3(\psi)} {\rm Im} \omega\ ; \label{eq:F2t0}\\
	& F_4 = J_\psi\left[ \frac12 F_0 - \mu \rho e^{-4A}\right] + J_\psi \wedge d\, {\rm Im} (\rho e^{-3A} v) \label{eq:F4t0}\ ;\\
	& F_6 = 0 \ .
	\end{align}
\end{subequations}
Unlike in section \ref{ssub:flux}, this time the flux equations for $F_4$ are not obviously following from the ones for $F_0$, or from any other combination of equations.

% subsection adst0 (end)

\subsection{Minkowski} % (fold)
\label{sub:mink}

The first equation in (\ref{eq:pspM}), $d_H \phi_+ = 0$, simply gives
\begin{equation}\label{eq:dJpsiM}
	\rho={\rm const}\ ,\qquad \theta={\rm const}\ ,\qquad d J_\psi=0 \ ,\qquad H=0 \ . 
\end{equation}
The second equation in (\ref{eq:pspM}), $d_H {\rm Re} \phi_-=0$, 
\begin{equation}\label{eq:revM}
	d (e^{-A}{\rm Re} v)=0 \ ,\qquad d {\rm Re} (i e^{-A} v\wedge \omega_\psi)= 0 \ .
\end{equation}
(The five-form part of $d_H {\rm Re} \phi_-=0$ can be shown to be redundant.)

The RR fluxes can now easily be computed from the third equation in 
\begin{align}
	F_0 &= - J_\psi^{-1}\llcorner d (\rho e^{-3A} {\rm Im} v)\ ;\\
	F_2 &= - J_\psi^{-1}\llcorner d {\rm Im} ( i \rho e^{-3A} v \wedge \omega_\psi)\ ;\\
	F_4 &= \frac12 F_0 \,J_\psi^2 + d ({\rm Im} \rho e^{-3A} v \wedge J_\psi)\ ; \\
	F_6 &=0\ .
\end{align}
Once again, the Bianchi identity for $F_4$ follows from the one for $F_0$, as in (\ref{eq:F4}), (\ref{eq:dF4}). 

% subsection mink (end)
% section manip (end)

\section{A general massive deformation} % (fold)
\label{sec:def}

Using the results of section \ref{sec:stt}, we will now point out the existence of a first-order AdS deformation of any SU(3) Minkowski solution in IIA. As we saw in the introduction, this includes any solution obtained as back-reaction of O6--D6 systems in IIA --- although in section \ref{sec:massiveo6} we will specialize it to the case of a single O6 in $\rr^6$. The expansion parameter is $\mu=\sqrt{-\Lambda/3}$. This deformation should not be taken as a modulus: as we will see below, the fluxes we will introduce contain $\mu$, and flux quantization will in general discretize it. Rather, our expansion is to be understood as a formal device to establish the existence of a solution at finite $\mu$.

We will start by determining how $\theta$ should be deformed. As we remarked after (\ref{eq:F2su3}), this parameter is an undetermined constant for the O6 solution we want to deform. However, we would like our solution to have something to do with the DGKT solution we reviewed in section \ref{sec:dgkt}. More specifically, we would expect our solution to approach the DGKT solution far from the source. Remembering (\ref{eq:th0}), we will take $\theta$ to be small. Since our deformation parameter is $\mu$, we might then take $\theta$ to be of order $\mu$.

This decision seems to run into trouble, however, as soon as we consider (\ref{eq:rev}). If $\theta$ is of order $\mu$, $v$ seems to diverge as $\mu \to 0$, whereas we need it to go to zero. 

To cure this potential disaster, we need at least two more factors of $\mu$ in the numerator of (\ref{eq:rev}). One 
can try to postulate that these extra factors are somehow supplemented by the derivative. This leads us to 
\begin{equation}\label{eq:deltatheta}
	\theta \sim \mu + \mu^3 \tau + \ldots\ .
\end{equation}
As in \cite{gaiotto-t-2}, we also suppose that everything is either odd or even in $\mu$, so that whatever function or form is already non-zero before the deformation will be unchanged at first order. 
This means, in particular, that we do not change the dilaton, internal metric and warping given in section \ref{sec:o6}. 
This gives
\begin{equation}\label{eq:rev0}
	{\rm Re} v = \frac \mu 2 e^A d\tau + O(\mu^2)\ .
\end{equation}

Also, since now $v$ is introduced at first order, we can mimic the procedure in \cite[Sec.~4.1]{gaiotto-t-2} and use it to deform an SU(3) structure into an \stt\ structure. 
 The conclusions reached in that reference can be summarized as follows. The function $\psi$ and the one-form $v$ start at first order: 
\begin{equation}
	\psi = \mu \,\psi_1 + O(\mu^2) \ ,\qquad
	v = \mu\, v_1 + O(\mu^2)\ ;
\end{equation}
the pure spinors have the form 
\begin{align}
	&\phi_+ = (1+ i \theta )e^{-iJ} + O(\mu^2) \ , \\
	&\phi_- = \left(\frac i \psi v \wedge \omega \right) + v 
	\wedge\left(1 + \frac 12 j^2 \right) + O(\mu^2)\ . 
\end{align}
Comparing the order $\mu^0$ part of $\phi_-$ with (\ref{eq:su3}), we get
\begin{equation}\label{eq:Ovo}
	\Omega= \frac i \psi_1 v_1 \wedge \omega \ ,
\end{equation}
which means, in particular, that $v_1$ is a $(1,0)$ form with respect to the almost complex structure defined by the three-form $\Omega$ of the SU(3) structure solution. This can be used to derive the imaginary part of $v_1$: 
\begin{equation}\label{eq:Iv0}
	{\rm Im} v_1 = \frac 12 e^A I\cdot d\tau \ ,\qquad
	v = \frac12 e^A \del \tau + O(\mu^2)\ ,
\end{equation}
where $I\cdot$ is the action of the almost complex structure determined by $\Omega$, and $\del$ is the corresponding Dolbeault operator.  
Finally, notice that (\ref{eq:Ovo}) can be inverted by writing 
\begin{equation}\label{eq:omega}
	\omega = -\frac i{2 \psi_1}\bar v_1 \,\llcorner \Omega\ .
\end{equation}

So far we have only looked at equation (\ref{eq:rev}) and to the algebraic constraints on the pure spinors $\phi_\pm$. We now turn to the other differential equations, starting with the ones that constrain the geometry. 

The first equation we consider is (\ref{eq:drho}), that at first order simply reads $d \rho =0$. In view of (\ref{eq:rho}), this is consistent with our postulate that $A$ and $\phi$ should not be deformed at first order. Comparing with (\ref{eq:rho0}), we see that $\rho_0$ is an odd function of $\mu$:
\begin{equation}
	\rho_0 = \frac1{g_s}\mu + O(\mu^3)\ .
\end{equation}
We have called the first coefficient in the expansion $1/g_s$, so as to conform with the value of $\rho$ in the particular solution (\ref{eq:F2O6}). 

Equation (\ref{eq:dJpsi}) is more problematic, because of the $\sin(\theta)$ in the denominator that makes the perturbation series  start at order $\mu^{-1}$ in the left-hand side. Enforcing again our policy that all our power series in $\mu$ be either even or odd function of $\mu$, we can expand $J_\psi$ up to second order:
\begin{equation}
	J_\psi = J + \mu^2 J_{(2)} + O(\mu^2)\ .
\end{equation}
Equation (\ref{eq:dJpsi}) is then, at order $\mu^{-1}$,
\begin{equation}
	dJ= 0 \ .
\end{equation}
This is one of the equations in the system we are deforming, as we can see from (\ref{eq:rho1}). At order $\mu$, (\ref{eq:dJpsi}) then gives
\begin{equation}\label{eq:dJ2}
	d\left[J_{(2)}+\left(\frac16-\tau\right) J\right] = 2 e^{-A}{\rm Re} \Omega \ .
\end{equation}
As we will see, this equation is the only one we will encounter in which $J_{(2)}$ appears at all, so at this order $J_{(2)}$ has nothing else to satisfy. The right hand side is automatically closed, because of (\ref{eq:F2su3}); but saying that it should be exact is a possible obstruction to deforming a given SU(3) structure Minkowski solution. 

We will now look at the fluxes. Our formula for $H$, (\ref{eq:H}), has a $\sin(\theta)$ in the denominator, just like (\ref{eq:dJpsi}). That would again force us to start our perturbation theory with negative powers of $\mu$. In this case, however, we can actually use (\ref{eq:dJpsi}) to rewrite $H$ so that it starts at first order: 
\begin{equation}\label{eq:h}
	H_{\rm phys} = \mu h + O(\mu^2) \ ,\qquad
	h = 2 {\rm Re} \Omega +d(\psi_1 {\rm Im} \omega)\ .
\end{equation}
Notice that the first term in $h$ is the same as the one for $H$ in the SU(3) structure solution given in (\ref{eq:SU3th0}), and the second term vanishes wherever $\psi_1$ tends to a constant. 

As for the RR fluxes, only $F_0$ and $F_4$ will be generated at first order; $F_2$ will keep the same expression it had at zeroth order, (\ref{eq:F2su3}). $F_0$ is given by
\begin{equation}\label{eq:f0}
	F_0 = \mu f_0 + O(\mu^3) \ ,  
	\qquad g_s f_0 = -J^{-1}\llcorner d ( e^{-\phi}{\rm Im} v_1) + 5  e^{-A-\phi}\ .
\end{equation}
We have expanded (\ref{eq:F0}) at first order in $\mu$, and used (\ref{eq:rho}).
As remarked after (\ref{eq:dF4}), that the Bianchi identity for $F_4$ follows from the one for $F_0$. So the only Bianchi identity we have to impose at first order is that
\begin{equation}\label{eq:F0fo}
	d f_0 = 0 \ .
\end{equation}
For completeness, however, we also give here the expression for $F_4$. Actually, the Laurent series for $F_4$ in (\ref{eq:F4}) starts with a term $\sim F_0 J^2/\mu^2$, which diverges like $\mu^{-1}$. So $F_4$ only becomes finite once one considers a finite $\mu$. This is not terribly worrying: as we anticipated at the beginning of this section, the expansion in $\mu$ is simply a formal device to establish the existence of a solution at finite $\mu$. In any case, the $\mu^{-1}$ terms disappear if we go back to the $\tilde F_k$, which are closed under $(d-H_{\rm phys}\wedge)$. We get
\begin{equation}\label{eq:f4}
	\tilde F_4 = \mu \tilde f_4 + O(\mu^3)\ ,\qquad
	 g_s\tilde f_4 = \left( \frac12 g_s f_0 - e^{-4A}\right) J^2 + J\wedge d(e^{-3A} {\rm Im} v_1) - \psi_1 {\rm Im} \omega \wedge J^{-1}\llcorner d(e^{-3A}{\rm Im} \Omega)\ .	
\end{equation}

Let us now summarize this section. We found a first-order perturbation of an SU(3) Minkowski solution which turns it into an AdS solution of \stt\ type. The perturbation parameter is $\mu=\sqrt{-\Lambda/3}$. The  only input is the function $\tau$ in (\ref{eq:deltatheta}), which has to satisfy (\ref{eq:F0fo}). One also has to solve (\ref{eq:dJ2}), but this simply requires to invert $d$. 

We are now going to apply this first-order deformation to O6 solutions.

% section def (end)

\section{Massive O6 solution} % (fold)
\label{sec:massiveo6}

In section \ref{sec:def}, we have found a procedure to deform any SU(3)-structure Minkowski solution at first order in $\mu= \sqrt{-\frac \Lambda3}$. In this section, we will try to promote this deformation to a fully-fledged supergravity solution.

Although the first-order deformation procedure can potentially be applied to any O6--D6 system, we will focus on the region around a single O6. This means that we will take the internal manifold to be $\rr^6$, with a single localized source as in (\ref{eq:deltaO6}). By doing this, we gain more symmetries than would be available for a general O6--D6 system; that will help us solve the system. 

However, as we anticipated in the introduction, this should not be understood too literally as a massive O6 ``in flat space''. Unlike for (\ref{eq:o6metric}), in the massive case the metric will not approach flat space far away from the source, simply because flat space is not a solution in the massive case. There are two new length scales associated with the massive problem, $\frac1{\mu}$ and $\frac1{g_s F_0}$, and the deviations from flat space asymptotics will become apparent at distances of the order of the smallest of these two length scales. The solution of this section should be thought of as a ``close-up'' around an O6 source in an ${\rm AdS}_4 \times M_6$ geometry where $M_6$ is \emph{compact} --- so the large $r$-behavior will not too important.

After some preliminaries in section \ref{sub:symmetries}, in section \ref{sub:first} we will specialize the general procedure of section \ref{sec:def} to a single O6. In section \ref{sub:higher} we will then promote it to a finite deformation; this will culminate in the numerical study of section \ref{ssub:num}, where we will find numerical solutions and describe their physical features, some of which were described in the introduction. We will also study the system at higher order in perturbation theory, in section \ref{sub:third}. In section \ref{sub:massivet0} we will show that choosing $\theta=0$ in the pure spinors (\ref{eq:nicepair}) does not lead to a solution. Finally, in section \ref{sub:massivemink} we will look briefly at the system for the Minkowski case; we also found numerical solutions in this case, but they do not seem to satisfy flux quantization. Moreover, we do not know of any Minkowski compactification that uses this ingredient. We will not describe these solutions in as much detail as the AdS ones.

\subsection{Symmetries} % (fold)
\label{sub:symmetries}

As in section \ref{sec:o6}, we will denote by $x^i$ the coordinates parallel to the O6, and by $y^i$ the coordinates transverse to it. 

The massless O6 solution is symmetric under rotations of the three $y^i$, rotations of the three $x^i$, and translations in the $x^i$: 
\begin{equation}\label{eq:isoso}
	{\rm ISO}(3)\times {\rm SO}(3)\ .
\end{equation}

It is already clear that the massive solution will not be symmetric under the whole group (\ref{eq:isoso}). As we have argued in section \ref{sec:dgkt}, we need to consider an \stt\ solution. One of the data in its definition is a complex one-form $v$; as we saw in section \ref{ssub:su3su3}, the algebraic constraints in (\ref{eq:stt}) demand in particular that $v\wedge \bar v \wedge j^2 \neq 0$ everywhere. So the real and imaginary part of $v$ are two linearly independent one-forms. However, the only linearly indepedent one-form which does not break any of the symmetries in (\ref{eq:isoso}) is
\begin{equation}
	dr= \frac1r y^i dy^i\ .
\end{equation}

Thus, in the massive solution the symmetry group (\ref{eq:isoso}) will be broken. In section \ref{sub:first}, we will see that a natural subgroup emerges when one applies the general first-order procedure of section \ref{sec:def} to the O6 solution of section \ref{sec:o6}. 

% subsection symmetries (end)

\subsection{First order deformation} % (fold)
\label{sub:first}

We will still demand that translation along the three internal coordinates $x^i$ parallel to the O6 should remain a symmetry. This will not be valid for a solution where there are several O6 sources, such as the one reviewed in section \ref{sec:dgkt}. However, this invariance will be restored when we get closer to an individual O6, which is the focus of the present paper.

Since everything can only depend on the transverse coordinates $y^i$, from now on we will use the notation
\begin{equation}
	\del_i \equiv \del_{y^i}\ .
\end{equation}
Using (\ref{eq:rev0}) and (\ref{eq:Iv0}), we then have
\begin{equation}\label{eq:vo6}
	v = -\frac i2 \mu Z^{-1/2} \del_i \tau (Z^{-1/4}dx^i +i Z^{1/4} dy^i)\ .
\end{equation}
Since $\tau$ depends on $r$ only, we have $\del_i = \frac{y^i}r \del_r$, and ${\rm Im} v$ is proportional to 
\begin{equation}\label{eq:ydx}
	y^i dx^i\ ,
\end{equation}
which breaks the symmetries (\ref{eq:isoso}) of the massless O6 solution, 
as anticipated in section \ref{sub:symmetries}. Indeed, the one-form (\ref{eq:ydx}) is neither invariant under either the SO(3) that rotates the transverse $y^i$, nor under the SO(3) that rotates the parallel $x^i$. It is still invariant, however, under the diagonal SO(3) that rotates both the $x^i$ and the $y^i$ simultaneously. Also, it is still invariant under translations along the $x^i$, as we stipulated at the beginning of this section. So (\ref{eq:vo6}) breaks (\ref{eq:isoso}) to 
\begin{equation}\label{eq:iso}
	{\rm ISO}(3)\ .
\end{equation}
It is not hard to list all the possible forms invariant under (\ref{eq:iso}); we have done so in appendix \ref{sec:forms}. We will see that the rest of the solution respects this smaller symmetry group.

Let us now go back to applying the first order procedure of section \ref{sec:def} to the O6 solution.\footnote{As remarked in section \ref{sec:o6}, we will deform one particular SU(3) structure which solves (\ref{eq:rho1}); for this reason, our massive solution will have only four supercharges, or ${\cal N}=1$ in four dimensions, just like the solutions in \cite{dewolfe-giryavets-kachru-taylor,acharya-benini-valandro}. Incidentally, it is easy to show that \emph{any} supersymmetric SU(3) structure solution with Romans mass has only four supercharges.} The next step is to impose (\ref{eq:F0fo}), namely that $F_0$, calculated at first order, is constant:  
\begin{equation}
	df_0=0 \ ,\qquad  g_s f_0 = -\frac12 \Delta \tau + 5 Z  = {\rm const.}\ ,
\end{equation}
where $\Delta\equiv \del_i \del_i$, and $g_s$ is the value of $e^\phi$ at infinity in the unperturbed solution (\ref{eq:F2O6}). Explicitly, using (\ref{eq:Z}), we get
\begin{equation}\label{eq:tau}
	\tau = \frac13 \left(  5- g_s f_0\right) r^2- 5 
	r_0 r\ ,
\end{equation}
setting to zero an inconsequential integration constant.

The other equation to be solved is (\ref{eq:dJ2}). This can be inverted to give
\begin{equation}\label{eq:J2exp}
	J_{(2)} = 
	-2 \left(\frac13 - \frac{r_0}{2r}+ \frac p{r^3}\right) \omega_{2,1} + 2 \omega_{2,2}
	-\frac{\alpha'}r \omega_{2,3} +\left(\tau-\frac16 + \alpha\right)\omega_{2,4}
\end{equation}
where a prime denotes $\del_r$. We have used the two-forms defined in (\ref{eq:2forms}); those forms are invariant under (\ref{eq:iso}), as promised. The constant $p$ and the function $\alpha=\alpha(r)$ are as yet undetermined.

At this point, we have already demonstrated the existence of a solution at first order. For completeness, however, let us also give the physical fluxes explicitly. First of all, we can determine $\psi_1$ from imposing that $J_\psi \to J$. Looking at the expression of $J_\psi$ in (\ref{eq:Jpsi}), this can be done by checking that $J^{-1}\llcorner \left(\frac i{2 \psi_1^2} v_1\wedge \bar{v_1}\right)=1$; we get 
\begin{equation}
	\psi_1 = \frac{\tau'}{2 \sqrt{Z}}\ .
\end{equation}
Now we can compute the first-order fluxes $\tilde f_4$ and $h$ from (\ref{eq:h}), (\ref{eq:f4}):
\begin{align}
	 g_s \tilde f_4 & = \frac 1{r^3}\left(-\frac52 r_0 + Z^{-1}\right) \omega_{4,1} +
	\left(\frac {r_0}{2r} - \frac13 (4 + g_s f_0) \right) \omega_{4,4}\ ;\\
	\nonumber
	h&= d\left[\left( -\frac{\tau' +2 r_0}{2 r} +\frac23 \right) \omega_{2,1} + \left(\frac{\tau'}{2rZ}-2\right) \omega_{2,2} +\frac12 \omega_{2,4}\right] \ .
\end{align}
As already stressed, the flux $F_2$ will not get deformed at first order in $\mu$.

Let us now pause to consider the properties of the first-order solution we have just obtained. First of all, we note that we have a certain freedom: we have left undetermined a function $\alpha(r)$ in (\ref{eq:J2exp}), which does not enter in the fluxes, and a constant $f_0$, defined in (\ref{eq:f0}) as the ratio between the deformation of $F_0$ and $\mu$. Let us see what happens if we set
\begin{equation}\label{eq:f05}
	 f_0=\frac 5{g_s}\ ,
\end{equation}
inspired by (\ref{eq:SU3th0}), which is valid for SU(3) structures.
We see that we cancel the $r^2$ term in (\ref{eq:tau}), which now goes linearly. One can then check that
\begin{equation}
	r \to \infty \quad \Rightarrow \quad f_4 \to \frac 32 J^2
	\ ,\qquad h\to 2 {\rm Re} \Omega \ ;
\end{equation}
in other words, far from the O6 source the solution approaches the SU(3) solution in (\ref{eq:SU3th0}). 

The perturbative procedure, however, can only work in an appropriate regime. We have already determined $J_{(2)}$ in (\ref{eq:J2exp}). Since $\tau$ actually grows with $r$, $J_{(2)}$ seems to grow large at large $r$, thus invalidating the first-order procedure. If $f_0=5/g_s$, for example, we see from (\ref{eq:tau}) that $\tau$ grows linearly; if $\alpha=0$, since $J_\psi=J+ \mu^2 J_{(2)}+ \ldots$, and recalling that $r_0= g_s l_s $, we have that the perturbation procedure is valid only if 
\begin{equation}
	r \ll  \frac 1 {g_s l_s \mu^2}\ .
\end{equation}
We are not necessarily interested, however, in what happens outside this region, because eventually we want to compactify the six ``internal'' directions, and in particular the three directions $y^i$. In the smeared solution we reviewed in section \ref{sec:dgkt}, we see from (\ref{eq:scales}) that 
the compactification radius in string units goes like $R\sim \mu^{-1/3}$, whereas $1/(g_s \mu^2)\sim \mu^{-3}$. In other words, the perturbative procedure breaks down for distances of order $\mu^{-3}$, which are much larger than the compactification radius $\mu^{-1/3}$.

In any case, we are now going to set up the study of the system of differential equations at all orders, guided by the results of this section. We will come back to perturbation theory in $\mu$ in section \ref{sub:third}.

% subsection first (end)

\subsection{Full solution} % (fold)
\label{sub:higher}

We now want to check whether the solution we just found at first order in $\mu$ survives beyond first order. We are not going to use perturbation theory in this section; we will go back to using it in section \ref{sub:third}.

\subsubsection{Variables} % (fold)
\label{ssub:variables}

At first order, the whole solution was determined by a single piece of data, the function $\tau$ in (\ref{eq:deltatheta}), which then has to solve (\ref{eq:F0fo}).

Beyond first order, however, the input data are many more: the functions $\psi$, $\theta$ and the forms $v$, $j$, $\omega$, in (\ref{eq:nicepair}), as well as the warping function $A$ in (\ref{eq:warped}). At first order, the continuous symmetry (\ref{eq:iso}) emerged, and we are going to assume that it is not broken in the full solution. This means that we should expand $v$ in terms of the one-forms (\ref{eq:1forms}), and $j$, $\omega$ in terms of the two-forms (\ref{eq:2forms}). 

There is also a discrete symmetry that we can use to our advantage. The solution we are looking for contains an O6, which is defined by quotienting the theory by the symmetry $\Omega (-)^{F_L} I_y$, where $\Omega$ is the world-sheet parity, $F_L$ is the fermionic number for left-movers, and 
\begin{equation}\label{eq:Iy}
	I_y: \left\{ \begin{array}{l}
		x^i \to x^i \\
		y^i \to -y^i
	\end{array}\right.
\end{equation}
 is the inversion in the three $y^i$ directions. The pure spinors $\phi_\pm$ should then transform as \cite{benmachiche-grimm}
\begin{equation}
	I_y^* \phi_+ = \lambda (\phi_+) \ ,\qquad
	I_y^* \phi_- = \lambda (\bar\phi_-)\ ,
\end{equation}
where $\lambda$ is the sign operator defined after (\ref{eq:stt}). This implies
\begin{equation}\label{eq:I*forms}
	I_y^* v = \bar v \ ,\qquad
	I_y^* j = -j \ ,\qquad
	I_y^* \omega = - \bar\omega\ .
\end{equation}
All the invariant forms in appendix \ref{sec:forms} transform by simply picking up a sign, as detailed in table \ref{parity}. Using that table, (\ref{eq:I*forms}) implies
\begin{equation}\label{eq:vjoO6}
	v= v_r \,\omega_{1,0} + i\, v_i \,\omega_{1,1}\ ,\qquad 
	j = \sum_{i=1}^4 j_i\, \omega_{2,i}
	\ ,\qquad
	\omega= a_0 \,\omega_{2,0} +i \sum_{i=1}^4 a_i\, \omega_{2,i}\ ;
\end{equation}
the coefficients $v_r$, $v_i$, $j_i$, $a_i$ are now all real.
% subsubsection variables (end)

\subsubsection{Algebraic equations} % (fold)
\label{ssub:alg}
With this parameterization in hand, we can now proceed to imposing the algebraic equations (\ref{eq:SU2}). These give:
\begin{subequations}\label{eq:jaO6}
		\begin{align}
			&j_4= j_3 r^2 \ ,\qquad 
			a_4= a_3 r^2 \ ,\qquad
			a_2 j_1 +a_1 j_2 = 2 a_3 j_3 r^2\ , \label{eq:jRoO6}\\
			&a_3^2 r^2 - a_1 a_2 = a_0^2=
				j_3^2 r^2 - j_1 j_2 \label{eq:j2O6}\ .
		\end{align}
\end{subequations}
Specifically, (\ref{eq:jRoO6}) comes from (\ref{eq:jRo}), whereas  (\ref{eq:j2O6}) comes from (\ref{eq:j2}). Moreover, the requirement in (\ref{eq:stt}) that $( \phi_-, \bar\phi_-) \neq 0$ demands\footnote{In fact, the first two equations in (\ref{eq:jRoO6}) are linear precisely because we divided by a common factor $a_0$, since it cannot vanish.} 
\begin{equation}
	a_0 \neq 0  \ .
\end{equation}

Given a solution to the algebraic constraints (\ref{eq:jaO6}), one can also compute the internal, six-dimensional metric associated to the pure spinors. This is not really needed in finding a solution, except for one important check: that its signature should be Euclidean. Applying (\ref{eq:metric}), we find 
\begin{equation}\label{eq:metricalfa}
	ds^2= (\alpha_1 \delta^{ij} + \alpha_2 y^i y^j) dx^i dx^j + (\alpha_3 \delta^{ij} + \alpha_4 y^i y^j) dy^i dy^j + \alpha_5 \epsilon_{ijk} y^i dx^j dy^k \ ,
\end{equation}
where the $\alpha_i= \alpha_i(r)$ are given by 
\begin{equation}\label{eq:aaj}
\begin{split}
		&\alpha_1 = \frac{-a_2 j_3 + a_3 j_2}{a_0} r^2 \ ,\qquad
		\alpha_2 = \frac{a_2 j_3 - a_3 j_2}{a_0} + \frac{v_i^2}{\tan^2(\psi)} \ , \\
		&\alpha_3 = \frac{a_1 j_3 - a_3 j_1}{a_0} r^2 \ ,\qquad
		\alpha_4 = \frac{-a_1 j_3 + a_3 j_1}{a_0} + \frac{v_r^2}{\tan^2(\psi)}
\end{split}
\qquad
\alpha_5 = \frac{a_2 j_1 - a_1 j_2}{a_0}\ .
\end{equation}
The metric (\ref{eq:metricalfa}) is symmetric under ISO(3), as we argued above (\ref{eq:iso}). If we go to polar coordinates for the $y^i$, by defining $r=\sqrt{y^i y^i}$ as in (\ref{eq:Z}), and 
\begin{equation}
	\hat y^i \equiv \frac {y^i} r \ ,
\end{equation}
we can write (\ref{eq:metricalfa}) as
\begin{equation}\label{eq:metricfibr}
\begin{split}
	ds^2= (\alpha_1 \delta^{ij} + r^2\alpha_2 \hat y^i \hat y^j) &Dx^i Dx^j + (\alpha_3 + r^2 \alpha_4) dr^2 + r^2 
	\left( \alpha_3 - \frac{r^2 \alpha_5^2}{4 \alpha_1}\right) ds^2_{S^2} \ ,\\ 
	&D x^i =dx^i - \frac{r^2\alpha_5}{2 \alpha_1} \epsilon^{ijk} \hat y^j d \hat y^k\ ,
\end{split}
\end{equation}
where $ds^2_{S^2}$ is the round metric of unit radius on the $S^2$ in the $y^i$ directions (which is the one that surrounds the O6). 
This exhibits the metric as a fibration of the $\rr^3$ spanned by the $x^i$ (along which the O6-plane is wrapped) over the $\rr^3$ spanned by the $y^i$, or by $r$ and the $\hat y^i$. Since the connection is a globally defined one-form, this fibration is topologically trivial. Notice that the function multiplying $dr^2$ simplifies to
\begin{equation}\label{eq:a34}
	\alpha_3 + r^2 \alpha_4 = 
	\left(\frac{r\, v_r}{\tan(\psi)}\right)^2\ ,
\end{equation}
using (\ref{eq:aaj}).

% subsubsection alg (end)

\subsubsection{Differential equations} % (fold)
\label{ssub:diff}
The differential equations we have to impose are (\ref{eq:rev}), (\ref{eq:dJpsi}), $dF_0=0$, and $dF_2=\delta_{\rm O6}$, where $F_0$ and $F_2$ are given by (\ref{eq:F0}) and (\ref{eq:F2}), and $\delta_{\rm O6}$ is given by (\ref{eq:deltaO6}). Recall that $dF_4=0$ follows from $dF_0=0$, as pointed out before (\ref{eq:dF4}). 

First of all, (\ref{eq:rev}) gives
\begin{equation}\label{eq:vr}
	v_r = - \frac{e^A}{2\mu r}\frac{\theta'}{\sin(\theta)}\ .
\end{equation}
(\ref{eq:dJpsi}) is clearly odd under $I_y$. From table \ref{parity}, we see that there are four odd three-forms; so (\ref{eq:dJpsi}) has four non-trivial components. One of these turns out to be algebraic:
\begin{equation}\label{eq:vi}
	v_i = \frac{e^A}{2\mu r^2}\frac{j_2}{a_0} \frac{\tan(\psi)}{\sin(\theta)}\ .
\end{equation}
So $v$ is completely determined algebraically, at all orders. The other three components in (\ref{eq:dJpsi}) are 
\begin{equation}\label{eq:dJpsico}
\begin{split}
	&\del_r\log\left(\frac{j_1 r^3}{\sin(\theta)\cos(\psi)}\right)=\frac{a_1}{j_1} \frac{\theta'}{\sin(\psi)}\ ,\\
	&\del_r\log\left(\frac{j_2 r}{\sin(\theta)\cos(\psi)}\right)=\frac{a_2}{j_2} \frac{\theta'}{\sin(\psi)}\ ,\\
	&\del_r\log\left(\frac{j_3 r^3}{\sin(\theta)\cos(\psi)}\right)=\left(a_3 - \frac{j_2 e^{2A}}{4 a_0 r^4 \mu^2}\frac{\cos^2(\psi)}{\sin^2(\theta)}\right)\frac{\theta'}{j_3\sin(\psi)}\ .
\end{split}
\end{equation}

We now turn to the Bianchi identities. We have one first-order equation that reads $F_0={\rm const}.$ After some manipulation we write it as an equation linear in the derivatives of the variables:
\begin{equation}\label{eq:bif0}
	\del_r \log \left(\frac {v_i r e^{-3A}}{\sin(\theta)}\right)=
	\theta' \cot^2(\psi) \left(\frac52 \cot(\theta) - \frac{F_0 e^{-4A}}{2 \mu \rho_0} + \frac{j_3 v_i \cos(\psi) e^A}{a_0^2 \mu \sin(\theta)}\right)\ .
\end{equation}
We also have $dF_2=\delta_{\rm O6}$. A priori, this would seem to have four components, since $F_2$ is odd under $I_y$. However, closer inspection reveals that only three components are non-trivial: 
\begin{equation}
	F_2 = \sum_{i=1}^4 f_{2,i} \omega_{2,i} \ ,\qquad
	dF_2 = (3 f_{2,1} + r f_{2,1}') \omega_{3,1} 
	+f_{2,2} \omega_{3,3} - \left(f_{2,3}+\frac1r f_{2,4}'\right) \omega_{3,5} +\frac1r f_{2,2}' \omega_{3,7}\ .
\end{equation}
The component of $dF_2$ along $\omega_{3,1}$ can be set to zero by taking $f_{2,1}$ proportional to $r^{-3}$; the proportionality constant can be fixed by requiring that it reproduces the correct factor in $\delta_{\rm O6}$. This can be read off (\ref{eq:F2O6}). Thus the non-trivial equations are three:
\begin{equation}\label{eq:fdiff}
	f_{2,1}= - \frac{l_s}{r^3} \ ,\qquad f_{2,2}=0 \ ,\qquad
	f_{2,4}'= -r f_{2,3}\ .
\end{equation}
These $f_{2,i}$ are determined by (\ref{eq:F2}) in terms of the data $j_i$, $a_i$, $\psi$, $\theta$, $A$ and their first derivatives. The equations for $f_{2,1}$ and $f_{2,2}$ give two equations which are again linear in the derivatives of the variables:
\begin{subequations}\label{eq:bif2}
	\begin{gather}
			\begin{split}
				&\del_r \log\left(\frac{a_1 v_i r^4 e^{-3A}}{\sin^2(\theta) \sin(2 \psi)}\right) = \frac{\theta'}{2 a_1}\left[  j_1 \left(-\frac 5{\sin(\psi)} +3 \sin(\psi)\right)\right.\\
				 +\frac{\cos^3(\psi)}{\sin(\psi)}&\left.\left(
				-\frac{l_s e^{4A}}{\rho_0 \mu r^3}\cos(\psi) - \frac{4 a_0^2}{j_2} - \frac{F_0 e^{4A} j_1 \cot(\theta)}{\rho_0 \mu}
				+ \frac{a_3 j_1 j_2 e^{2A}}{a_0^2 \mu^2 r^2 \sin^2(\theta)}
				\right)\right]\ ,
			\end{split}\\
		\begin{split}
			&\del_r \log\left(\frac{a_2 v_i r^2 e^{-3A}}{\sin^2(\theta)\sin(2 \psi)}\right) = \\
			&\frac{j_2 \theta'}{2 a_2}\left[-2 \sin(\psi)+\frac{\cos^3(\psi)}{\sin(\psi)}\left(-5- \frac{F_0 e^{4A}  \cot(\theta)}{\rho_0 \mu} + \frac{a_3 j_2^2 e^{2A}}{a_0^2 r^2 \mu^2\sin^2(\theta)}  \right)\right]\ .
		\end{split}
	\end{gather}	
\end{subequations}
Remarkably, by using these two equations and (\ref{eq:dJpsico}), one can show that the last equation in (\ref{eq:fdiff}) is actually automatically satisfied. 

All in all, we have three differential equations from (\ref{eq:dJpsico}) (coming from (\ref{eq:dJpsi})), one from (\ref{eq:bif0}) (coming from  $F_0=$const), and two from (\ref{eq:bif2}) (coming from $dF_2 = \delta_{\rm O6}$), for a total of six. All of these are first-order, and linear in the first derivatives. 

Having counted our equations, let us now count our variables. We can use (\ref{eq:vr}) and (\ref{eq:vi}) to eliminate $v_r$ and $v_i$ from the system; moreover, we can use the first two in (\ref{eq:jRoO6}) to eliminate $j_4$ and $a_4$. It is less clear how to use the remaining three equations in (\ref{eq:jaO6}); one possibility is to derive $a_1$, $j_1$ and $j_3$. This leaves us with the variables
\begin{equation}\label{eq:var}
	a_0 \ ,\qquad a_2 \ ,\qquad a_3 \ ;\qquad
	j_2 \ ;\qquad
	A \ ,\qquad \theta \ ,\qquad \psi\ , 
\end{equation}
for a total of seven variables. We should also notice, however, that we have not yet fixed the gauge invariance coming from reparameterizations of the radial direction: 
\begin{equation}\label{eq:rr}
	r \to \tilde r(r)\ .
\end{equation}
Under these reparameterizations, the coefficients of $j$ and $\omega$ a priori could mix. It turns out, however, that only the coefficients of $\omega_3$ and $\omega_4$ mix; if we impose the algebraic equations in (\ref{eq:jaO6}), even the coefficients along those two are proportional. So, in particular we have
\begin{equation}
	a_0 \to \left(\frac r{\tilde r}\right)^2 a_0 \ ,\qquad
	(a_2,j_2) \to \left(\frac r{\tilde r}\right) (a_2,j_2) \ ,\qquad
	a_3 \to \left(\frac r{\tilde r}\right)^3 a_3 \ ,
\end{equation}
whereas of course $A$, $\theta$, $\psi$ transform as functions. 

Thus, out of the seven variables in (\ref{eq:var}), one is redundant because of the gauge invariance (\ref{eq:rr}). This effectively leaves us with six variables, which is as many as the differential equations (\ref{eq:dJpsico}), (\ref{eq:bif0}), (\ref{eq:bif2}). So we have as many equations as variables, and we expect a solution to exist. We will now study the system numerically.

% subsubsection diff (end)

\subsubsection{Numerics} % (fold)
\label{ssub:num}

The system we found in section \ref{ssub:diff} is first-order, and linear in the derivatives of our variables. We found it useful to fix the gauge invariance (\ref{eq:rr}) by demanding $\theta$ to be exact at order $\mu^3$; namely, 
\begin{equation}\label{eq:thetagf}
	\theta_{\rm gf}= \mu + \mu^3 \tau\ ,
\end{equation}
with $\tau$ given (\ref{eq:tau}). In other words, the $\ldots$ terms in (\ref{eq:deltatheta}) are absent. This gauge makes it easier to compare the massless limit of our numerical solutions with the solution in section \ref{sec:o6}.

Also, we imposed boundary conditions at an $r$ much larger than $r_0=g_s l_s$, but much smaller than the scales $(g_s F_0)^{-1}$ and $\mu^{-1}$, where deviations from the massless asymptotics become apparent. Using the first-order solution in section \ref{sub:first} as a clue, we identified a family of boundary conditions (depending on $F_0$ and $\mu$) such that, when one takes the limit $F_0\to 0$ and $\mu\to0$ (thus forgetting for a moment about flux quantization), one recovers the massless solution\footnote{The family is obtained with the help of the perturbative expansion we will consider in section \ref{sub:third}; actually, besides $F_0$ and $\mu$, the family also depends on an integration constant in $a_3$. This constant has no influence on the massless limit.}. This works quite well, especially if one takes the limit by keeping $\frac{g_s F_0}{\mu}=5$, as in the special choice (\ref{eq:f05}) for the first-order solution. We take all this as a check that our numerical analysis is sound.

\begin{figure}[t]
	\centering
	\subfigure[The massless O6 solution.]
	{
		\includegraphics[scale=.55]{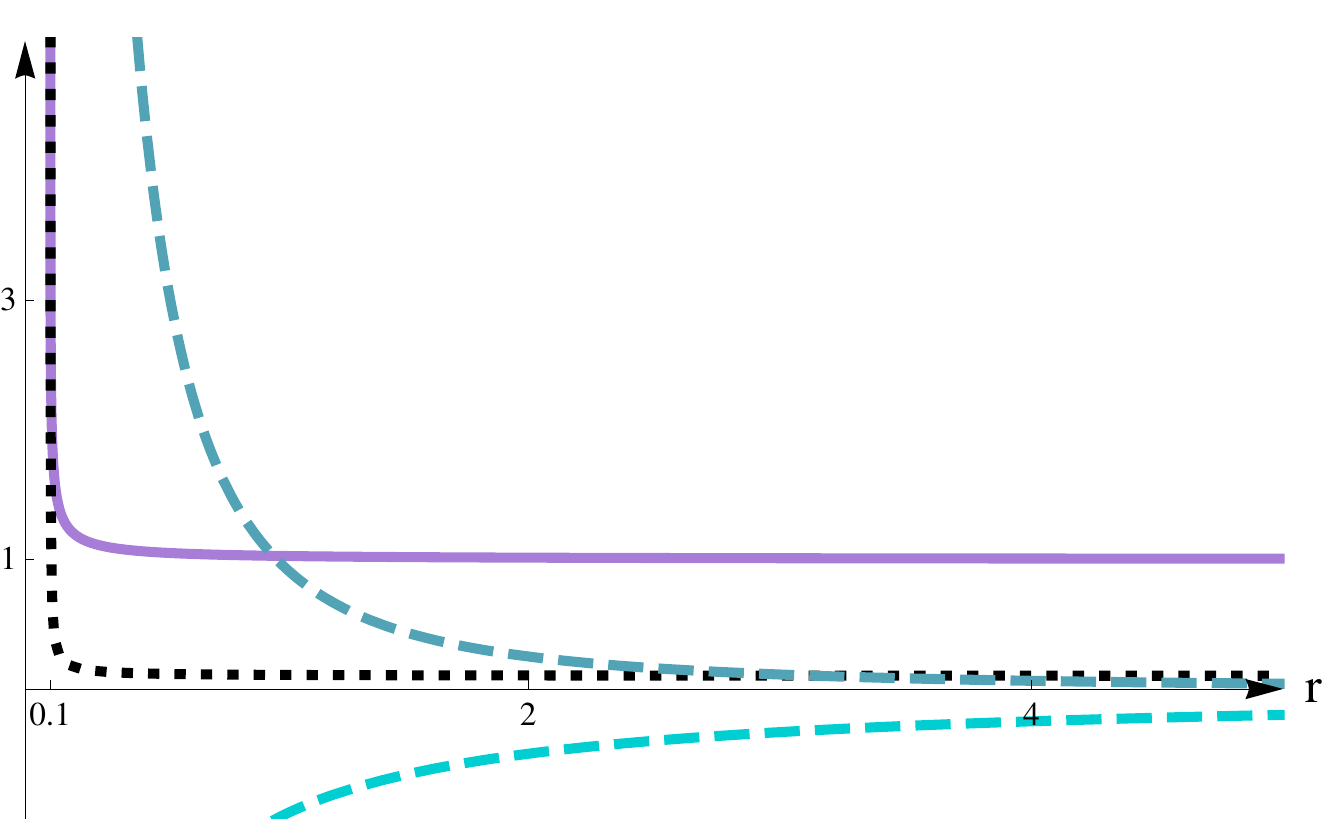}
		\label{fig:o60}
	}
	\subfigure[A O6 solution with Romans mass.]
	{
		\includegraphics[scale=.55]{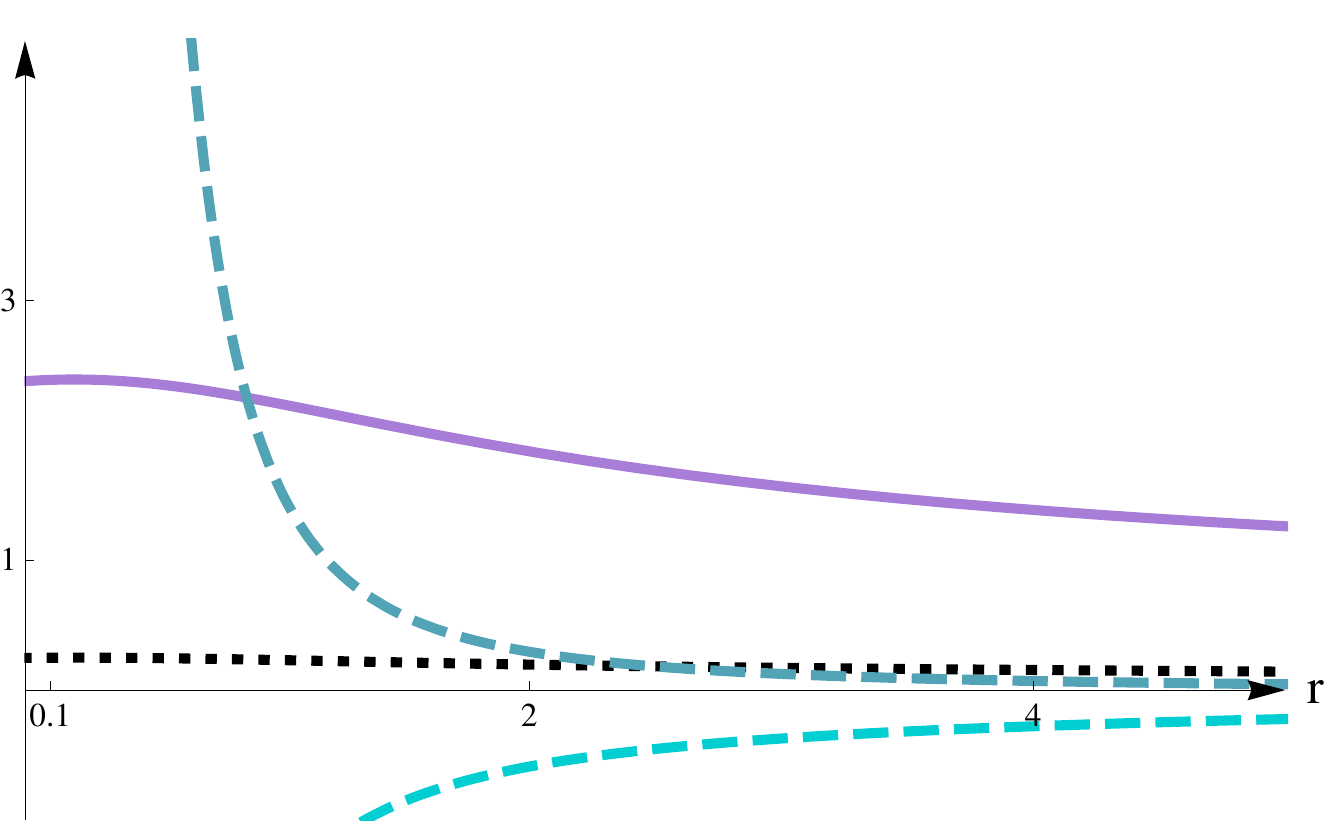}
		\label{fig:o6m}
	}
	\caption{\small Comparison between the massless O6 solution and a solution with Romans mass. The solid line is $e^A$; the dotted line is $e^\phi$; the dashed lines are $j_3$ (positive) and $a_0$ (negative). On the left we plot these coefficients (in string units, for $g_s=0.1$) for the solution with $F_0=0$: from (\ref{eq:o6metric}) and (\ref{eq:JOO6}) we get $e^A=(1-r_0/r)^{-1/4}$, $j_3=1/r^2$, $a_0=-1/r$. In particular, the solution diverges at $r=r_0=0.1\,l_s$. On the right, we plot the same coefficients for a supersymmetric solution with localized O6 source, for $\mu\sim .055$, $F_0=\frac{4}{2\pi l_s}$. $j_3$ and $a_0$ retain a power-law behavior, while $e^A$ no longer diverges at $r_0=0.1$. At larger distances, one can see deviations from the flat-space behavior, due to the fact that flat space is not a solution for $F_0\neq0$, as observed earlier.}
	\label{fig:o6num}
\end{figure}

We then increased $F_0$ until it satisfied the flux quantization condition $F_0 = \frac{n_0}{2\pi}$, $n_0 \in \zz$. The behavior of the solutions for $n_0 \neq 0$ is qualitatively different from the massless solution: notably, it does not display the divergence at $r_0=g_s l_s$ that plagues the massless solution (\ref{eq:o6metric}) --- see figure \ref{fig:o6num}. 
We checked that the eigenvalues of the metric (\ref{eq:metricalfa}) remain positive in our numerical solutions. 

Let us now focus on the asymptotic behavior of our solutions at $r\to 0$. In our gauge, $\theta$ tends to a constant at $r\to 0$; numerically, one can see $\psi$ and $A$ also tend to constants $\psi_0$ and $A_0$. We can then use the differential equations (\ref{eq:dJpsico}), (\ref{eq:bif0}), (\ref{eq:bif2}) to find the asymptotic behavior of the coefficients $a_i$, $j_i$:
\begin{equation}\label{eq:ajas}
	\begin{split}
		a_0 \sim a_{00} r^{-2}&\ ,\qquad a_1 \sim a_{10} r^{-3} \ ,\qquad
		a_2 \sim a_{20} r^{-1}\ ,\qquad a_3 \sim a_{30} r^{-3}\ ; 
		\\
		&j_1 \sim j_{10} r^{-3}\ ,\qquad j_2 \sim j_{20} r^{-1} \ ,\qquad j_3 \sim j_{30} r^{-3}\ , 
	\end{split}
\end{equation}
where the $a_{i0}$ and $j_{i0}$ are constants. These are also in agreement with the algebraic constraints (\ref{eq:jaO6}).

From (\ref{eq:ajas}) it follows that the $\alpha_i$ in (\ref{eq:metricalfa}) behave as
\begin{equation}
	\alpha_1 \to \alpha_{10} \ ,\qquad
	\alpha_2 \sim \alpha_{20} r^{-2} \ ,\qquad
	\alpha_3 \sim \alpha_{30} r^{-2} \ ,\qquad
	\alpha_4 \sim \alpha_{40} r^{-4} \ ,\qquad
	\alpha_5 \sim \alpha_{50} r^{-2}\ ,
\end{equation}
where $\alpha_{i0}$ are non-zero constants. For the crucial combination $\alpha_3 + r^2 \alpha_4$, however, which multiplies $dr^2$ in (\ref{eq:metricfibr}), from (\ref{eq:a34}) and (\ref{eq:vr}) we see that
\begin{equation}
	\alpha_3 + r^2 \alpha_4 \to \left(\frac52 g_s \mu\right)^2\ ;
\end{equation}
thus, the $r^{-2}$ divergencies cancel out, and this coefficient goes to a constant.

As $r \to 0$, the metric (\ref{eq:metricfibr}) then tends to 
\begin{equation}
	\begin{split}
		ds^2= (\alpha_{10} \delta^{ij} +\alpha_{20} \hat y^i \hat y^j) &D_0x^i D_0x^j + \left(\frac52 g_s \mu\right)^2 dr^2 + 
		\left( \alpha_{30} - \frac{\alpha_{50}^2}{4 \alpha_{10}}\right) ds^2_{S^2} \ ,\\ 
		&D_0 x^i =dx^i - \frac{\alpha_{50}}{2 \alpha_{10}} \epsilon^{ijk} \hat y^j d \hat y^k\ .
	\end{split}
\end{equation} 
This metric factorizes in a factor $dr^2$, and a five-dimensional $\rr^3$ fibration over $S^2$. Thus, asymptotically we have $\rr\times M_5$. 

For most values of $\mu$, the curvature of $M_5$ is small, and we can trust the supergravity approximation. However, the size of the $S^2$ remains finite, and the metric is no longer geodesically complete. Fortunately, it is possible to perform an analytic continuation by going to polar coordinates for the $y^i$. One can then see that, in the system described in sections \ref{ssub:alg} and \ref{ssub:diff}, all explicit dependence on $r$ drops out; the only dependence is introduced by the way we fix the gauge freedom (\ref{eq:rr}). One can then continue $r$  to negative values. With our gauge choice (\ref{eq:thetagf}), one can see that for $r<0$ the metric gets continued essentially to a mirror copy of itself.  
  
One might feel unsatisfied by the fact that the $S^2$ that surrounds the orientifold never shrinks to a zero size; so the O6-plane locus does not really exist in these metrics, even though all fields transform as they should under the antipodal map $\hat y^i \to - \hat y^i$ of an O6 projection. Even in the massless case, however, the transverse $S^2$ does not shrink in the smooth Atiyah--Hitchin metric (see for example the discussion in \cite[Sec.~3]{gibbons-manton-AH}).

For special choices of $\mu$, the curvature of $M_5$ gets large; in that case, the supergravity approximation breaks down. It is possible that $\alpha'$ corrections make the size of the $S^2$ shrink, but this is of course speculation.

% subsubsection num (end)

% subsection higher (end)

\subsection{Back to perturbation theory in $\mu$} % (fold)
\label{sub:third}

In section \ref{sub:first} we considered our equations to order $\mu$, and found an explicit solution. In section \ref{sub:higher} we analyzed the conditions for unbroken supersymmetry in the setup suggested by the first-order solution, culminating in the numerical analysis in \ref{ssub:num}. In this section we will go back to perturbation theory in $\mu=\sqrt{-\frac \Lambda3}$, to see how explicit can the solution be made.   

First, a bit of notation: we are going to expand the various coefficients and functions as a power series in $\mu$, keeping the same assumptions in section \ref{sec:def} about which expansions contain  even or odd powers:
\begin{equation}\label{eq:exp3}
	\begin{split}
		j_i= j_{i,0} + \mu^2 j_{i,2} &+ \mu^4 j_{i,4}+ O(\mu^4) \ ,\qquad
		 \psi= \mu \psi_1  + \mu^3 \psi_3 + O(\mu^5)\ ,\\
		a_i= a_{i,0} + \mu^2 a_{i,2} &+\mu^4 a_{i,4}+ O(\mu^4)\ ,\qquad
		A= A_0 + \mu^2 A_2 + O(\mu^4)\ ,\\
		&\theta= \mu + \mu^3 \tau +  \mu^5 \theta_5 + O(\mu^7) \ .
	\end{split}
\end{equation}
As it turns out, the equations at order $\mu^2$ and $\mu^3$ mix quite a bit. Using the algebraic equations, we found it convenient to use the variables
\begin{equation}\label{eq:var3}
	A_2 \ ,\qquad \theta_5 \ ,\qquad \psi_3 \ ;\qquad
	 j_{1,4}\ ,\qquad j_{2,4}\ ,\qquad j_{3,2} \ ,\qquad a_{2,2}\ .
\end{equation}
For example, even if we have already solved $J_{(2)}$ at second order in (\ref{eq:J2exp}), we did so only up to an unknown function $\alpha(r)$. This means that one component was actually undetermined; in terms of the expansion (\ref{eq:exp3}), this remaining equation can be written in terms of the variables (\ref{eq:var3}). At the same time, of the equations in (\ref{eq:dJpsi}) only two contain the variables in (\ref{eq:var3}); the third involves variables at higher order, and we can ignore it at this level. We then have one equation $F_0=\,$const.~and three equations from $dF_2= \delta_{\rm O6}$, just like in our discussion at all orders in section \ref{ssub:diff}. 

In section \ref{ssub:diff}, the system  of differential equations was first-order and linear in the derivatives of the variables. The perturbative system we are considering in this subsection, once we use the solution found at first order in section \ref{sub:first}, is also linear (inhomogenous) in the variables themselves. This means that we can write it as
\begin{equation}
	v'= M v + b \ ,\qquad
	v=(\theta_5,\psi_3,j_{1,4},j_{2,4},j_{3,2},a_{2,2})^t \ .		
\end{equation}
The matrix $M$ is particularly simple in the gauge  $A=A_0=\log(Z^{-3/4})$, and with the simplifying assumption $f_0=5$:
\begin{equation}\label{eq:Msys}
	M=
	\left(\begin{array}{cccccc}
		0 & -2\sqrt{Z} & 0 & 0 & 10 r_0 r^2 & \tau Z\\
		0 & \frac12\left(\frac3r -\frac Z{r_0}\right) & 0 & 0 &-\frac{\tau}Z & -\frac52 r_0 \\
		0 & 0 & \frac3r & 0 & 8 r Z & -4Z^{3/2}\\
		0 & 0 & 0 & \frac1r & -4 r & 0 \\
		0 & 0 & 0 &-\frac1{2r^3} & \frac2r & \frac {\sqrt{Z}}{r^2}\\
		0 & 0 & 0 &-\frac1{r^2\sqrt{Z}} & \frac2{\sqrt{Z}} &  \frac12\left(\frac3r -\frac Z{r_0}\right)
	\end{array}\right)\ .
\end{equation}
The expression for the vector $b$ is more complicated, and we see no reason to inflict it on the reader. The first three columns of (\ref{eq:Msys}) show three obvious eigenvalues; the variables $\theta_5$, $\psi_3$, $j_{1,4}$ are determined once the other three are. So the crucial part of $M$ is the lower-right $3\times 3$ block, concerning the variables $j_{2,4}$, $j_{3,2}$, $a_{2,2}$. The eigenvalues of this block can be found by the Cardano--Tartaglia formula, and so in principle the system at this order can be solved analytically. 

% subsection third (end)

\subsection{The special case $\theta=0$} % (fold)
\label{sub:massivet0}

In section \ref{sec:stt} we have divided the analysis of \stt\ structure solutions in three cases: AdS for $\theta \neq 0$ (the ``generic'' case of section \ref{sub:ads}), AdS for $\theta=0$ (the ``special'' case of section \ref{sub:adst0}), and the Minkowski case (in section \ref{sub:mink}). So far, in this section we have analyzed the system in detail in the generic AdS case $\theta\neq0$. We now want to go back to the other two cases. We will begin in this subsection by the special AdS case, $\theta=0$. 

We will again work with the symmetry group (\ref{eq:iso}), for the same reasons explained in section \ref{sub:symmetries} and \ref{sub:first}. The parameterization of the forms $v$, $j$, $\omega$ is still the same as in section \ref{ssub:variables}. The algebraic equations satisfied by them can still be written as in (\ref{eq:jaO6}).

Since in this case $H$ in (\ref{eq:Ht0}) is not already exact (as for (\ref{eq:h})), we have to impose by hand that $dH=0$. Since $H$ is odd, the only non-zero component of this equation is the one along $\omega_{4,0}$: 
\begin{equation}
	a_2 \, v_r \, \mu = 0 \ .
\end{equation}
$v_r$ cannot be zero because of the requirement $(\phi_-,\bar \phi_-)=0$ in (\ref{eq:stt}). Also, $\mu\neq0$ by assumption; so we get
$a_2=0$.

We then look at $d(\rho J_\psi)=0$, again from (\ref{eq:Ht0}). This has four non-zero components, but in particular the one along $\omega_{3,3}$ tells us that
\begin{equation}
	j_2 = 0 \ .
\end{equation}

We can now go back to the algebraic system (\ref{eq:jaO6}), and use that $a_2=j_2=0$. The last equation of (\ref{eq:jRoO6}) tells us that $a_3 j_3=0$. But, both if $a_3=0$ and if $j_3=0$, (\ref{eq:j2O6}) now tells us $a_0=0$. This means that ${\rm Re} \omega=0$, which is not possible, again because of the requirement $(\phi_-, \bar \phi_-)\neq 0$ in (\ref{eq:stt}).

Thus, in this section we have quickly disposed of the case $\theta =0$. This case cannot lead to massive O6 solutions with the symmetry (\ref{eq:iso}). 

% subsection massivet0 (end)

\subsection{Minkowski} % (fold)
\label{sub:massivemink}

Finally, in this section we will look at the Minkowski case. 

Once again, we can use the parameterization of the forms $v$, $j$, $\omega$  in section \ref{ssub:variables}, whose coefficients have to satisfy the algebraic equations in (\ref{eq:jaO6}). 

The relevant differential equations were given in \ref{sub:mink}. We start with (\ref{eq:dJpsiM}). This says 
\begin{equation}\label{eq:dJpsicoM}
	j_2=0 \ ,\qquad \frac{r^3 j_1}{\cos(\psi)}= {\rm const}. \ ,\qquad
	\del_r\log\left(\frac{r^3 j_3}{\cos(\psi)}\right) = -\frac{v_r v_i}{ r j_3} \frac{\cos^3(\psi)}{\sin^2(\psi)} \ .
\end{equation}
We then turn to (\ref{eq:revM}). The first is trivially satisfied, using the symmetries of our setup. The second gives
\begin{equation}\label{eq:wreq2}
	\del_r \log\left(\frac{a_0 v_i r^3 e^{-A}}{\sin(\psi)}\right)= \frac{a_2}{a_0}\frac{v_r}{v_i \cos(\psi)}\ .
\end{equation}

We now turn to the Bianchi identities. They can be discussed along the lines of the AdS case in section \ref{ssub:diff}. One consists in imposing that $F_0$ is constant, and can be written as
\begin{equation}\label{eq:bif0M}
	\del_r\log( v_i r e^{-3A} ) = \frac{v_r r}{\tan^2(\psi)}\left( \frac{2 j_3 v_i \cos(\psi)}{a_0^2}- F_0 e^{3A} \right) \ .
\end{equation}
As in (\ref{eq:fdiff}), $F_2$ would seem to give three equations. The ones for $f_{2,1}$ and $f_{2,2}$ read: 
\begin{subequations}\label{eq:bif2M}
	\begin{align}
		&\del_r \log \left( \frac{a_1 v_i r^4 e^{-3A}}{\sin(2 \psi)}\right)= \frac{v_r}{a_1} \cos(\psi) \left(-\frac{2 a_0}{v_i r} + \frac{2 a_3 j_1 v_i r}{a_0^2 \tan^2(\psi)} - \frac{l_s\cos^2(\psi) e^{3A}}{\sin(\psi)r^2}\right)\\
		&\del_r \log \left( \frac{a_2 v_i r^2 e^{-3A}}{\sin(2 \psi)} \right)= 2\frac{\cos^3(\psi)}{\sin^2(\psi)}\frac{a_3 j_2 v_r v_i r}{a_2 a_0^2}
	\end{align}
\end{subequations}
Once again, the third equation in (\ref{eq:fdiff}) can be shown to be automatically implied by (\ref{eq:bif2M}) and by (\ref{eq:dJpsicoM}), (\ref{eq:wreq2}). 

So we have one differential equation from (\ref{eq:dJpsicoM}), one from (\ref{eq:wreq2}), one from (\ref{eq:bif0M}), and two from (\ref{eq:bif2M}). This gives a total of five differential equations, which are all first order, and linear in the derivatives. 

Let us now count our variables. Unlike in the AdS case, $v_r$ and $v_i$ are now independent variables. On the other hand, (\ref{eq:dJpsicoM}) allows us to eliminate $j_2$ (which vanishes) and $j_1$ (which is a function of other variables). All in all, we can take as independent variables
\begin{equation}\label{eq:varM}
	a_3 \ ,\qquad j_3 \ ,\qquad v_r \ ,\qquad v_i \ ,\qquad
	A \ ,\qquad \psi \ . 
\end{equation}
Just as in section \ref{ssub:diff}, we still have the gauge freedom (\ref{eq:rr}). This means that one of these six variables is actually redundant, and we effectively have five variables. 

So we again have as many variables as equations. We have studied the system numerically. The solutions share some qualitative features with the ones for the AdS case (see figure \ref{fig:o6m}); for example, the warping $A$ stays flat rather than diverging. However, they only survive for small values of $F_0$, which do not satisfy the flux quantization condition $F_0 = \frac {n_0}{2\pi l_s}$. For values of $F_0$ that do satisfy flux quantization, the system seems to crash in a singularity before it gets to $r=0$.

It is also possible to set up a perturbative study. Since $\Lambda=0$ in this case, we cannot perturb in $\mu$. We introduce a new perturbation parameter $\nu$, such that $v\to 0$ as $\nu \to 0$. This can be achieved by taking the coefficients $v_r$ and $v_i$ to be odd functions of $\nu$, while the other coefficients $a_i$, $j_i$ will be even functions of $\nu$. We solved the resulting system at first order in $\nu$, similarly to section \ref{sub:first}. 

Finally, it would presumably also be possible to deform the Minkowski solutions discussed in this section into an AdS solution, by generalizing the procedure in section \ref{sec:def}. 

% subsection massivemink (end)

% section massiveo6 (end)

\section*{Acknowledgments.}
Our research is supported in part by MIUR--PRIN contract 2009-KHZKRX. We would like to thank G.~Moore for his initial involvement in the project, and A.~Zaffaroni for useful discussions.

\appendix

\section{Forms} % (fold)
\label{sec:forms}

In this appendix, we will give a basis of forms symmetric under the symmetry ISO(3) we identified in section \ref{sub:first}. This consists of translations in the directions $x^i$ parallel to the O6-plane, and of simultaneous rotations of both the $x^i$ and of the $y^i$, transverse to the O6-plane. In the main text, we have used this basis to expand both our pure spinors and fluxes.  

The one-forms are:
\begin{subequations}\label{eq:1forms}
	\begin{align}
	\omega_{1,0}&=y^i\,dy^i\equiv r\,dr\ ,\\
	\omega_{1,1}&=y^i\,dx^i\ .
	\end{align}
\end{subequations}
A 2-form basis compatible with the symmetry is:
\begin{subequations}\label{eq:2forms}
	\begin{align}
	\omega_{2,0}&=\epsilon_{ijk}\,y^i\,dy^j\wedge dx^k\ ,\\
	\omega_{2,1}&=\epsilon_{ijk}\,y^i\,dy^j\wedge dy^k\ ,\\
	\omega_{2,2}&=\epsilon_{ijk}\,y^i\,dx^j\wedge dx^k\ ,\\
	\omega_{2,3}&=y^i\,dy^i\wedge y^j\,dx^j= \omega_{1,0}\wedge \omega_{1,1} \ ,\\
	\omega_{2,4}&=dx^i\wedge dy^i=J\ ;
	\end{align}
\end{subequations}
we recalled here that the last form is nothing but the two-form $J$ of the massless O6 solution, (\ref{eq:JOO6}).

The 3-forms can be written in terms of:
\begin{subequations}\label{eq:3forms}
	\begin{align}
	\omega_{3,0}&=\dfrac{1}{6}\epsilon_{ijk}\,dx^i\wedge\,dx^j\wedge dx^k\equiv {\rm vol}_{\parallel}\ ,\\
	\omega_{3,1}&=\dfrac{1}{6}\epsilon_{ijk}\,dy^i\wedge\,dy^j\wedge dy^k\equiv {\rm vol}_{\perp}\ ,\\
	\omega_{3,2}&=\epsilon_{ijk}\,dx^i\wedge\,dy^j\wedge dy^k\ ,\\
	\omega_{3,3}&=\epsilon_{ijk}\,dx^i\wedge\,dx^j\wedge dy^k\ ,\\
	\omega_{3,4}&=\epsilon_{ijk}y^i\,y^m\,dx^m\wedge\,dy^j\wedge dy^k=\omega_{1,1}\wedge\omega_{2,2}\ ,\\
	\omega_{3,5}&=y^i\,dx^j\wedge dy^i\wedge dy^j=\omega_{1,1}\wedge\omega_{2,4}\ ,\\
	\omega_{3,6}&=y^i\,dx^j\wedge dx^i\wedge dy^j=\omega_{1,0}\wedge\omega_{2,4}\ ,\\
	\omega_{3,7}&=\epsilon_{ijk}\,y^i\,r\,dr\wedge dx^j\wedge dx^k=\epsilon_{ijk}\,y^i\,y^m\,dy^m\wedge dx^j\wedge dx^k\ ,
	\end{align}
\end{subequations}	
4-forms and 5-forms can then be obtained as wedge products from the previous definitions:
\begin{subequations}\label{eq:4forms}	
	\begin{align}
	\omega_{4,0}&=\epsilon_{ijk}y^i\,dx^m\wedge\,dx^j\wedge dy^m\wedge dy^k=\omega_{2,0}\wedge\omega_{2,4}\ ,\\
	\omega_{4,1}&=\epsilon_{ijk}\epsilon_{lmn}\,y^i\,y^l\,dx^j\wedge dx^m\wedge dy^k\wedge dy^{n}=\omega_{2,1}\wedge\omega_{2,2}\ ,\\
	\omega_{4,2}&=y^i\,dx^i\wedge {\rm vol}_{\perp}=\omega_{1,1}\wedge {\rm vol}_{\perp}\ ,\\
	\omega_{4,3}&= {\rm vol}_{\parallel}\wedge y^i\,dy^i= {\rm vol}_{\parallel}\wedge \omega_{1,0}\ ,\\
	\omega_{4,4}&=dx^i\wedge dx^j\wedge dy^i\wedge dy^j=-J^2\ ;
	\end{align}
\end{subequations}
\begin{subequations}\label{eq:5forms}
	\begin{align}
	\omega_{5,0}&=\omega_{2,2}\wedge {\rm vol}_{\perp}\ ,\\
	\omega_{5,1}&=\omega_{2,1}\wedge {\rm vol}_{\parallel}\ .
	\end{align}
\end{subequations}
Crucially, this basis is closed under exterior derivative $d$ wedge product. One can then express both in terms of appropriate tensors: for example, the wedge product between the 2-form $\Psi=\Psi_{i}\omega_{2,i}\,, (i=0, \dots, 4)$ and the 3-form $\Omega=\Omega_{I}\omega_{3,I}\,, (I=0, \dots, 7)$ can be written in terms of a tensor $W_{23}$: 
\begin{equation}
\Psi\wedge\Omega=\Psi_{i}\,\Omega_{I}\,\omega_{2,i}\wedge\omega_{3,I}\equiv \Psi_{i}\, \Omega_{I}\,(W_{23})_{i, I, \alpha}\omega_{5,\alpha}=(\Psi\wedge\Omega)_{\alpha}\omega_{5,\alpha} \ ,
\end{equation}
where $\alpha=0,1$.
The same idea can be applied to the exterior derivative. For example:
\begin{equation}
d\,\Psi=d(\Psi_{i}\,\omega_{2,i})=\dfrac{\Psi_{i}'}{r}\omega_{1,0}\wedge\omega_{2,i}+\Psi_{i} d(\omega_{2,i})\equiv\left[\dfrac{\Psi_{i}'}{r}(W_{12})_{0, i, I}+\Psi_{i} D_{i, I}\right] \omega_{3,I}\ ,
\end{equation}
with $D_{i,I}$ an appropriate tensor. Working out all these tensors speeds up computations significantly. 

Under the a parity transformation 
\begin{equation}
\sigma:y_{i}\rightarrow -y_{i}
\end{equation}
in the directions perpendicular to the O6-plane, the forms defined above transform by picking up signs. This signs are summarized in table \ref{parity}. 
\begin{table}[ht]\caption{Parity properties of our form basis under $I_y$ in (\ref{eq:Iy}).}\label{parity}
\begin{center}
 \begin{tabular}{c|c|c|}\cline{2-3}
&Even & Odd\\ 
\hline\hline
\multicolumn{1}{|c||}{1-forms} & $\omega_{1,0}$ &$\omega_{1,1}$\\
\hline
\multicolumn{1}{|c||}{2-forms}& $\omega_{2,0}$ &$\omega_{2,1},\omega_{2,2},\omega_{2,3},\omega_{2,4}$\\
\hline
\multicolumn{1}{|c||}{3-forms} & $\omega_{3,0},\omega_{3,2},\omega_{3,4},\omega_{3,6}$ & $\omega_{3,1},\omega_{3,3},\omega_{3,5},\omega_{3,7}$\\
\hline
\multicolumn{1}{|c||}{4-forms} & $\omega_{4,1},\omega_{4,2},\omega_{4,3},\omega_{4,4}$ &$\omega_{4,0}$\\
\hline
\multicolumn{1}{|c||}{5-forms} & $\omega_{5,0}$ &$\omega_{5,1}$\\
\hline
\end{tabular}
\end{center}
\end{table}

% section forms (end)

%% bibliography generated by
%\bibliography{at}
%\bibliographystyle{at}
%% and then cut-and-pasted before submitting
\providecommand{\href}[2]{#2}

\end{document}